\documentclass{aa}  

\usepackage{txfonts}
\usepackage[dvips]{graphicx}

\begin{document}
   \title{MHD simulations of jet acceleration from Keplerian accretion disks}

   \subtitle{The effects of disk resistivity}

   \author{C. Zanni
          \inst{1,2}
          \and
          A. Ferrari
          \inst{2,3}          
	  \and
	  R. Rosner
	  \inst{3}
	  \and
	  G. Bodo
      \inst{4}
      \and	  
      S. Massaglia
	  \inst{2}
          }

   \offprints{C. Zanni}

   \institute{Laboratoire d'Astrophysique de Grenoble, 414 Rue de la Piscine, 
              BP 53, F-38041 Grenoble, France\\
              \email{Claudio.Zanni@obs.ujf-grenoble.fr}
         \and
              Dipartimento di Fisica Generale dell'Universit\`a, via Pietro Giuria 1, 
              10125 Torino, Italy\\
             \email{ferrari@ph.unito.it}
	 \and
	      Department of Astronomy and Astrophysics, University of Chicago, 
	      5640 S. Ellis Av., Chicago, IL 60637, USA\\
	      \email{r-rosner@uchicago.edu} 
         \and
              Osservatorio Astronomico di Torino, Viale Osservatorio 20, 
              10025 Pino Torinese, Italy\\
              \email{bodo@to.astro.it}
             }

   \date{Received ..................... ; accepted ................... }

 
  \abstract
   {Accretion disks and astrophysical jets are used to model many active astrophysical objects, viz., 
young stars, relativistic stars, and active galactic nuclei. However, extant proposals on how these 
structures may transfer angular momentum and energy from disks to jets through viscous 
or magnetic torques do not yet provide a full understanding of the physical mechanisms involved. 
Thus, global stationary solutions do not permit an understanding of the stability of these structures; and 
global numerical simulations that include both the disk and jet physics have so far been limited to 
relatively short time scales and small (and possibly astrophysically unlikely) ranges of viscosity and 
resistivity parameters that may be crucial to define the coupling of the inflow-outflow dynamics. }
   {Along these lines we present in this paper self-consistent time-dependent simulations of supersonic jets
launched from magnetized accretion disks, using high resolution numerical techniques. In particular
we study the effects of the disk magnetic resistivity, parametrized through an $\alpha$-prescription, in determining
the properties of the inflow-outflow system. Moreover we analyze under which conditions steady state solutions of the type 
proposed in the self-similar models of Blandford \& Payne can be reached and maintained in a self-consistent nonlinear stage.}
   {We use the resistive MHD FLASH code with adaptive mesh refinement, allowing us to follow the evolution 
of the structure for a time scale long enough to reach steady state. A detailed analysis of the initial configuration 
state is given.}
   {We obtain the expected solutions in the axisymmetric (2.5 D) limit. Assuming a magnetic field around equipartition with
    the thermal pressure of the disk, we show how the characteristics of the disk-jet system, as the ejection efficiency and the energetics,
    are affected by the anomalous resistivity acting inside the disk.} 
    {}

   \keywords{Accretion disks --
             Jets and outflows --
             Magnetohydrodynamics --
             Methods: numerical
              }

   \maketitle
%
%

\section{Introduction}
\label{sec:intro}
Astrophysical jets are an important component in many active astrophysical objects, 
from young stellar objects (YSO) to relativistic stars and galactic nuclei.
In particular, Herbig-Haro (HH) outflows 
are detected in stellar forming regions around T-Tauri stars, 
characterized mainly by optical emission line spectra (e.g. Reipurth \&
Bally \cite{Reipurth}). On a much larger scale, relativistic radio jets
are accelerated in the innermost cores of Active Galactic Nuclei (AGN)
(e.g. Giovannini \cite{Giovannini}): their emitting component is synchrotron 
relativistic electrons, with a cold proton component or, most likely, a Poynting 
flux electromagnetic component (De Young \cite{DeYoung}).

Despite of being characterized by extremely different space, time and energy scales,
it is commonly accepted that all these systems derive their energy from accretion onto 
a central object (Livio \cite{Livio}): the physical origin of these supersonic outflows
has been related to the dynamical evolution of magnetized accretion disks around 
a deep gravitational well. 
Even if the acceleration and collimation mechanisms of jets are still not clear, some basic models
have been proposed and shown to be successful (for rather recent reviews see Pudritz et al. \cite{Pudritz} 
for YSO and Ferrari \cite{Ferrari}, \cite{Ferrari04} for AGN). The
overall idea is to extract energy and angular momentum from the
accreting matter and to feed it into a plasma that is accelerated in
two opposite directions along the rotation axis of the disk. Lovelace
(\cite{Lovelace}) and Blandford (\cite{Blandford}) proposed
independently that this can be done by electromagnetic forces. In
particular Blandford \& Payne (\cite{BlPayne82}) derived a steady
state MHD solution for an axisymmetric magnetocentrifugally driven
outflow from a Keplerian disk; for the outflow to be launched the
poloidal magnetic field lines must be inclined less than $60^0$ with
respect to the plane of the accretion disk.
On the other hand, Sauty \& Tsinganos (\cite{Sauty1994}), Sauty et al. (\cite{Sauty2002}),
Sauty et al. (\cite{Sauty2004})  have derived meridionally self-similar models to study 
the launching mechanism from the hot corona of the central object.

The magnetocentrifugal mechanism has been the subject of a
series of numerical studies (Ustyugova et al. \cite{Ustyugova}, Ouyed \& Pudritz \cite{Ouyed}, 
Krasnopolsky et al. \cite{Krasno}, Anderson et al. \cite{Anders04}, Fendt \cite{Fendt}, Pudritz \cite{Pudritz2}) 
based on ideal MHD simulations in which the disk is treated as a boundary condition.
On one hand they show how a steady solution can be obtained in
a few dynamical time scales and how the acceleration, collimation and 
stationarity of the outflow depend on the mass loading from the disk and 
on the magnetic field structure. On the other hand  the back reaction of the
outflow on the disk can not be taken into account.

When the structure of the magnetized accretion disk is included self-consistently 
in the models, a diffusive mechanism must be introduced inside the disk to balance 
the shearing due to differential rotation and the inward advection of field lines. 
Moreover also viscous torques, which can transport angular momentum radially inside
the disk itself, should be, in principle, taken into account.

For instance, K\"onigl (\cite{Konigl}), Wardle \& K\"onigl (\cite{WarKon}) and Li (\cite{Li1996}) 
studied the structure of an ambipolar diffusion dominated disk connecting it with a 
Blandford \& Payne solution at the disk surface. Ogilvie \& Livio (\cite{Ogilvie}) solved the 
vertical structure of a thin (optically thick) magnetized disk taking into account an effective $\alpha$
turbulent viscosity and resistivity.

In a series of papers including turbulent $\alpha$ resistivity (Ferreira \& Pelletier \cite{FerPel1995}, 
Ferreira \cite{Ferreira1997}), viscosity (Casse \& Ferreira \cite{CasFe00a}) and entropy generation 
(Casse \& Ferreira \cite{CasFe00b}) the authors
calculated radially self-similar stationary solutions of accretion-ejection structures.
Two important features emerged clearly from these latter models: first of all it was shown
that, in order to balance the magnetic and gravitational compression on the disk itself, 
the thermal energy must be around equipartition with the magnetic energy inside the disk.
With these conditions, the vertical thermal pressure gradient is the only force that can
push the mass at the surface of the disk to be accelerated in the outflow. Second,
the magnetic torque must change sign at the disk surface, in order to extract angular momentum
from the disk and transfer it to the outflow. This condition determines a strong constraint on the
resistive configuration: the magnetic diffusivity must be rather high ($\alpha \sim 1$) and anisotropic, 
the diffusion of the poloidal field being smaller than the diffusion of the toroidal component.

The first time-dependent numerical simulations in which the structure of the magnetized accretion 
disks is included (Uchida \& Shibata \cite{UchiShi} or more recently Kato et al. \cite{Kato}) showed that the interaction 
between a  geometrically thin rotating disk and a large scale magnetic field that was initially uniform and vertical
originates a transient state in which a strong toroidal field is generated that expels matter in the 
direction perpendicular to the disk plane (``sweeping magnetic twist mechanism''). 
One of the limits of these models is that the short simulated time scales and the ideal MHD approximation lead to
the formation of a transient and highly unstable outflow. Moreover the small density contrast between the disk and the surrounding
corona assumed in Uchida \& Shibata (\cite{UchiShi}) enables the disk to lose its angular momentum on a very short time scale.
These works have been recently updated in Kuwabara et al. (\cite{Kuwabara})  who studied the evolution of a thick magnetized torus 
assuming a constant resistivity throughout the computational domain: even if the presence of such a high and uniform resistivity must
be justified, the authors showed a quasi-stationary outflow launched from the in the inner radii of the torus.

Up to the present, the best effort to produce an accretion-ejection structure recurring to time-dependent
simulations has been performed by Casse \& Keppens (\cite{CasKe02}, \cite{CasKe04}) who showed 
how a quasi-stationary jet can be launched from the equipartition regions of a resistive accretion disk.
On the other hand, their resistive configuration, isotropic with $\alpha=0.1$ is rather different from the one predicted by
the self-similar steady models (Casse \& Ferreira \cite{CasFe00a}): despite of the use a stretched grid, 
the resolution of these simulations is rather low and it is likely that numerical dissipative effects are quite important.

In the present paper we present a numerical study of resistive MHD
axisymmetric accretion-ejection structures performed with the high
resolution code FLASH using Adaptive Mesh Refinement. The aim of the
paper is to simulate the disk-jet configuration over long time scales
in order to determine the effects of different configurations of an $\alpha$ 
resistivity, by varying its value and its degree of anisotropy, in determining
the properties of the system. Moreover we want to test whether a stationary 
state corresponding to the Blandford \& Payne self-similar solution can be 
reached and maintained. 
A critical point in the simulation is of course the choice
of the initial configuration, in particular the magnetic configuration
and the equilibrium of the disk. We have solved for an analytic self-similar 
equilibrium configuration of the disk including gravitational, centrifugal, thermal 
pressure and Lorentz forces; this solution defines also the initial magnetic
field.  We do not include physical viscosity: as it has been shown both in stationary
(Casse \& Ferreira \cite{CasFe00a}) and time-dependent contexts (Meliani et al. \cite{Meliani}), 
for conventional values of the disk turbulence ($\alpha \sim 0.1 - 1$, Prandtl number $\sim 1$) the
viscous torque is less efficient than the magnetic in extracting angular momentum
from the disk.

The paper is organized as follows. Section 2 is dedicated to
illustrating the equations and the numerical code used; the initial
configuration and the boundary conditions are discussed in detail. In
Section 3 the numerical simulations are presented for different
magnetic diffusivities and anisotropies. These parameters are linked
to the mass accretion rate, required torque and liberated accretion
energy. The acceleration mechanisms at the disk boundary and far above
the disk are discussed, also referring to the established current
circuits. The jet ejection efficiency is discussed in Section 4 in
relation to the accretion rates. In Section 5 we discuss the angular
momentum transport and in Section 6 the energy budget of the
inflow/outflow system is evaluated. 
In Section 7 we test if one of our cases can be characterized as 
a stationary solution.
Comments about the effect of Ohmic dissipation in the disk are
given in Section 8. Finally, Section 9 summarizes our results
and puts them in context with other analytical and numerical models.


\section{The numerical model}
\subsection{MHD equations}
\label{sec:equations}
We model the interaction between an accretion disk and the magnetic
field which threads it within a resistive MHD framework. The system of
equations that we solve numerically conveys therefore the conservation
of mass:
\begin{equation}
\label{eq:cmass}
\frac{\partial \rho}{\partial t}+\nabla\cdot(\rho \vec {u}) = 0 \; ,
\end{equation}
where $\rho$ is the mass density and $\vec{u}$ is the flow speed; the momentum equation: 
\begin{equation}
\label{eq:momentum}
\frac{\partial \rho\vec{u}}{\partial t} + \nabla \cdot \left[ 
       \rho \vec{u}\vec{u} +
       \left( P + \frac{\vec{B}\cdot\vec{B}}{2} \right)\vec{I}-
       \vec{B}\vec{B}
       \right] + \rho \nabla\Phi_\mathrm{g} = 0 \; ,
\end{equation}
where $P$ is the thermal pressure and $\vec{B}$ is the magnetic field. This equation takes into
account the action of thermal pressure gradients, Lorentz forces and gravity as determined by the
potential $\Phi_\mathrm{g} = -GM/\sqrt{r^2+z^2}$ representative of the gravitational field of a central object of mass
$M$.
The evolution of the magnetic field is determined by the induction equation (Faraday's law):
\begin{equation}
\label{eq:induction}
\frac{\partial \vec{B}}{\partial t} + \nabla \times \vec{E} = 0 \; ,
\end{equation}
where the electric field $\vec{E}$ is determined by the Ohm's law $\vec{E} = -\vec{u}\times\vec{B} + \bar{\bar{\vec{\eta}}} \vec{J}$.
Notice that the equations are written in a non-dimensional form, hence without $4\pi$ and $\mu_0$ coefficients.
The electric current $\vec{J}$ appearing in the Ohm's relation is determined by the Ampere's law $\vec{J} = \nabla\times\vec{B}$.
The magnetic resistivity $\bar{\bar{\vec{\eta}}}$ is indicated as a two-tensor to take into account anisotropic diffusive effects:
in our simulations we will consider a diagonal resistivity tensor $\eta_{ij}$ whose non-zero components are $\eta_{\phi\phi} = \eta_\mathrm{m}$
and $\eta_{rr} = \eta_{zz} = \eta_\mathrm{m}'$.

Finally the conservation of energy is expressed by
\begin{equation}
\label{eq:cenergy}
\frac{\partial e}{\partial t} + \nabla\cdot\left[
    \left(e + P + \frac{\vec{B}\cdot\vec{B}}{2}\right)\vec{u}-
    \left(\vec{u}\cdot\vec{B}\right)\vec{B} +
    \bar{\bar{\vec{\eta}}}\vec{J}\times\vec{B}
    \right] = -\Lambda_\mathrm{cool} \; ,
\end{equation}
where the total energy density 
\[
e = \frac{P}{\gamma-1}+\frac{\rho\vec{u}\cdot\vec{u}}{2}+\frac{\vec{B}\cdot\vec{B}}{2} + \rho\Phi_\mathrm{g}
\]
is given by the sum of thermal, kinetic, magnetic and gravitational energy. $\gamma = 5/3$ is the polytropic index of the
gas. $\Lambda_\mathrm{cool}$ is a cooling term defined by the parameter $0<f<1$:
\begin{equation}
f = \frac{\Lambda_\mathrm{cool}}{\Lambda_\mathrm{diss}}
\label{eq:cool}
\end{equation}
given by the ratio between the specific radiated energy and the Ohmic heating term $\Lambda_\mathrm{diss} = 
\bar{\bar{\vec{\eta}}}\vec{J}\cdot\vec{J}$. The parameter $f$ therefore determines the fraction of magnetic energy which is radiated away
instead of being dissipated locally inside the disk increasing its entropy.
Finally the system of equations is closed by the equation of state of ideal gases $P = n K T$ where $n = \rho/m_\mathrm{p}$ 
($m_\mathrm{p}$ being the proton mass) is the number density of the gas, $T$ is its temperature and $K$ is the Boltzmann constant.

To solve the resistive MHD system of Eqs. (\ref{eq:cmass}-\ref{eq:cenergy}) we employ a modified version of the MHD module
provided with the public code FLASH\footnote{FLASH is freely available at http://flash.uchicago.edu} (Fryxell et al. \cite{Fryxell}) developed 
at the ASC FLASH Center at the University of Chicago, adopting its Adaptive Mesh Refinement (AMR) capabilities. The simulations have
been carried out in 2.5 dimensions, that is in cylindrical geometry in the coordinates $r$, $z$ assuming axisymmetry around the rotation axis
of the disk-jet system. The algorithm implemented belongs to the class of high resolution Godunov schemes which are the best suited to study
supersonic flows: we therefore used a linear reconstruction of primitive variables with a minmod limiter on pressure and flow speed and a Van Leer 
limiter on density and on the magnetic field components; second order accuracy in time is obtained thanks to an Hancock predictor step on the primitive 
variables while, to compute the fluxes needed to update the conservative variables, we implemented an HLLE solver, that is
an approximate linearized Riemann solver, that assumes a-priori a two-wave configuration for the solution. 
To control the solenoidality of the magnetic field ($\nabla\cdot\vec{B}=0$) the eight-wave approach (Powell et al. \cite{Powell}) was used,
which is known for simply advecting the monopoles, while a parabolic diffusion operator (Marder \cite{Marder}, see also Dedner et al. \cite{Dedner}) 
was added to the induction equation to diffuse them.  We finally ensured to use an angular momentum conserving form of the $\phi$ component 
of the momentum equation Eq. (\ref{eq:momentum}) and a poloidal current flux conserving form of the $\phi$ component of the induction equation Eq. (\ref{eq:induction}).

\subsection{Initial conditions}
\label{sec:incond}
In the initial setup of our simulations we model a disk rotating with a slightly sub-Keplerian speed threaded by an initially purely poloidal magnetic
field. The disk initial model is derived imposing equilibrium between the forces initially intervening inside the disk, namely, gravity, centrifugal force, thermal 
pressure gradients, and Lorentz force; the disk setup is therefore a solution of the following system of equations:
\begin{eqnarray}
\label{eq:equilz}
\frac{\partial P}{\partial z} & = &  -\rho\frac{\partial\Phi_\mathrm{g}}{\partial z}-J_\phi B_r \\
\label{eq:equilr}
\frac{\partial P}{\partial r}  & = &  -\rho\frac{\partial\Phi_\mathrm{g}}{\partial r} +J_\phi B_z+ \frac{\rho u_\phi^2}{r}
\end{eqnarray}

The system of Eqs. (\ref{eq:equilz}-\ref{eq:equilr}) can be easily solved in separable variables assuming radial self-similarity; with this assumption 
all the physical quantities $U$ are given by the product of a power law for $r$ with a function of the variable $x=z/r$
\begin{equation}
\label{eq:selfsim}
U = U_0 \left(\frac{r}{r_0}\right)^{\beta_U}f_U\left(\frac{z}{r}\right) \; ,
\end{equation} 
where $z=0$ corresponds to the midplane of the disk. $f_U$ is an even function such that $f_U(0) = 1$ or an odd function such that $f_U(0) = 0$, 
depending on the symmetry of the variable with respect to the midplane of the disk. For the physical quantities which have an even symmetry 
($P,\; \rho,\; u_r,\; u_\phi,\; B_z$) $U_0$ therefore represents their value at $r = r_0,\; z=0$.

Self-similarity requires that all the characteristic speeds, namely sound and Alfven speed,  and flow speeds should scale
as the Keplerian velocity ($\propto r^{-1/2}$) on the midplane of the disk. Imposing also a polytropic relation between the disk density and pressure, 
that is $P = P_0\left(\rho/\rho_0\right)^\gamma$, the power law coefficients $\beta_U$ are determined as follows:
\begin{displaymath}
\begin{array}{ll}
\beta_{u_\phi} = \beta_{u_r} = \beta_{u_z} = -1/2 & \qquad \beta_P = -5/2 \\
\beta_{B_r} = \beta_{B_z} = -5/4                        & \qquad \beta_\rho = -3/2  \; .
\end{array}
\end{displaymath}

The initial poloidal magnetic field has been set through the flux function $\Psi$ to ensure the solenoidality of the field:
\begin{equation}
\Psi = \frac{4}{3}B_{z0}r_0^2\left(\frac{r}{r_0}\right)^{3/4}\frac{m^{5/4}}{\left(m^2+z^2/r^2\right)^{5/8}}
\end{equation}
The components of the field, obtained thanks to the simple relations:
\[
B_z = \frac{1}{r}\frac{\partial \Psi}{\partial r} \qquad B_r  = -\frac{1}{r}\frac{\partial \Psi}{\partial z}
\]
obviously fulfill the self-similarity requirements. The parameter $m$ determines the height scale on which the initial magnetic field bends, where
a value $m \to \infty$ gives a perfectly vertical ($B_r = 0$) field.

Given the poloidal magnetic field, the vertical equilibrium Eq. (\ref{eq:equilz}) can be numerically solved to give the vertical profiles of disk density ($f_\rho$) 
and pressure ($f_P = f_\rho^\gamma$). On the other hand the radial equilibrium Eq. (\ref{eq:equilr}) can be solved to determine the disk rotation speed
$u_\phi$.  The solution will therefore depend on the following non-dimensional parameters:
\begin{equation}
\epsilon = \left. \frac{c_\mathrm{s}}{V_\mathrm{K}}\right|_{z=0} = \left. \sqrt{\frac{p}{\rho}\frac{r}{GM}} \; \right|_{z=0}
\end{equation}
which is the ratio between the sound speed $c_\mathrm{s} = \sqrt{P/\rho}$ and the Keplerian rotation speed $V_\mathrm{K}=\sqrt{GM/r}$ evaluated 
on the disk midplane: this quantity determines the disk thermal height scale $H$ through the relation $H=\epsilon r$ (see Frank et al. \cite{Frank}, 
Ferreira \& Pelletier \cite{FerPel1995}); the magnetization parameter:
\begin{equation}
\mu = \left. \frac{B^2}{2P}\right|_{z=0}
\end{equation}
which gives the ratio between the magnetic and thermal pressure evaluated on the disk midplane. 
The disk rotation velocity at the midplane is slightly sub-Keplerian, with the deviation depending on the parameters ($\epsilon$, $\mu$ and $m$):
\begin{equation}
\left. u_{\phi}\right|_{z=0} = \left[1-\frac{5}{2}\epsilon^2 -2\epsilon^2\mu\left(\frac{5}{4}+\frac{5}{3m^2} \right)\right]\sqrt{\frac{GM}{r}}
\end{equation}
The toroidal velocity decreases towards the disk surface and it falls down to zero where the radial Lorentz force balances the gravitational pull and
the thermal pressure gradient.

For the components of the magnetic diffusivity tensor $\bar{\bar{\vec{\eta}}}$ acting in the disk we adopt an $\alpha$ prescription 
(Shakura \& Sunyaev \cite{Shsu}) in the same vein  of Ferreira (\cite{Ferreira1997} and related works) and Casse \& Keppens 
(\cite{CasKe02}, \cite{CasKe04}). 
Despite of being a common way to parametrize the transport coefficients inside an accretion disk, the origin of this
anomalous diffusivity is still debated. One of the most promising hypothesis states that the anomalous transport is of turbulent origin, 
triggered by some disk instability, being the magneto-rotational instability (MRI, Balbus \& Hawley \cite{BalHal}) the most accredited one.
The $\eta_{\phi\phi} = \eta_\mathrm{m}$ component is parametrized as follows:
\begin{equation}
\eta_\mathrm{m} = \alpha_\mathrm{m} \left. V_\mathrm{A}\right|_{z=0} H \exp \left( -2\frac{z^2}{H^2}\right) \; ,
\label{eq:mdiff}
\end{equation}
where $\alpha_\mathrm{m}$ is a constant parameter, $\left. V_\mathrm{A}\right|_{z=0} = \left. \left(B_z/\sqrt{\rho}\right)\right|_{z=0}$ is the Alfven speed 
calculated on the disk midplane and $H = \left. \left(c_\mathrm{s}/\Omega_\mathrm{K}\right)\right|_{z=0}$ is the thermal height scale of the disk. In the
simulations both the Alfven speed and $H$ are allowed to evolve in time.
The other components of the diffusivity tensor $\eta_{rr} =\eta_{zz} = \eta_\mathrm{m}'$ are assumed to be proportional to $\eta_\mathrm{m}$ through an 
anisotropy parameter $\chi_\mathrm{m}$, which is the inverse of the analogous parameter introduced by Ferreira \& Pelletier (\cite{FerPel1995}):
\begin{equation}
\chi_\mathrm{m} = \frac{\eta_\mathrm{m}'}{\eta_\mathrm{m}}
\end{equation}
A ratio $\chi_\mathrm{m} = 1$ indicates an isotropic resistive configuration.
We recall that the presence of an effective resistivity inside the disk allows the magnetic field to break the ``frozen-in'' condition and the matter to slip through
the field lines. The component $\eta_\mathrm{m}$, therefore indicated as \emph{poloidal} resistivity, allows the flow to slip through the field in the poloidal plane 
while $\eta_\mathrm{m}'$, indicated as \emph{toroidal} resistivity, controls the diffusion of the toroidal component of the field.

In a steady situation the accretion motion should be consistent with the poloidal magnetic field configuration as stated by the poloidal induction equation 
in its stationary form:
\begin{equation}
\label{eq:addif}
u_zB_r - u_rB_z = \eta_\mathrm{m}J_\phi 
\end{equation}
which expresses the balance between the advection and the diffusion of the poloidal magnetic field. Therefore we initially impose an accretion flow inside the
disk solving Eq. (\ref{eq:addif}) with the condition $u_z = \frac{z}{r}u_r$. Due to the self-similarity requirements, the radial accretion speed scales initially as 
the Keplerian speed on the midplane of the disk:
\begin{equation}
\left. u_{r}\right|_{z=0} = -\alpha_\mathrm{m}\left(2\mu\right)^{1/2}\epsilon^2\left(\frac{5}{4}+\frac{5}{3m^2} \right) \sqrt{\frac{GM}{r}}
\end{equation}
As the poloidal resistivity $\eta_\mathrm{m}$ exponentially decreases towards the disk surface, the initial accretion flow cancels out outside the disk.

It must be pointed out that in the initial conditions, since there is no toroidal magnetic field, there is no mechanism of angular momentum transport 
which can support the initial accretion flow: the angular momentum transport associated with the toroidal field will be triggered by torsional Alfven waves
due to the differential rotation between the midplane and the surface of the disk. Moreover, since we will assume a magnetic field around equipartition 
with the thermal energy on the midplane of the disk (see Section \ref{sec:simulations}), the time scale on which the transport of angular momentum will become
effective, given by the Alfven crossing time of the disk thickness, is comparable to the local period of rotation of the Keplerian disk and to its epicyclic frequency.

On top of the disk we prescribe an hydrostatic spherically symmetric atmosphere for which we impose, as for the disk, a polytropic relation between
pressure $P_\mathrm{a}$ and density $\rho_\mathrm{a}$: $P_\mathrm{a} = P_{\mathrm{a}0} \left( \rho_\mathrm{a}/\rho_{\mathrm{a}0}\right)^\gamma$.
The density and pressure distribution will be therefore given by:
\begin{equation}
\rho_\mathrm{a} = \rho_{\mathrm{a}0}\left( \frac{r_0}{R} \right)^{\frac{1}{\gamma-1}} \qquad 
P_\mathrm{a} = \rho_{\mathrm{a}0}\frac{\gamma -1}{\gamma}\frac{GM}{r_0}\left( \frac{r_0}{R} \right)^{\frac{\gamma}{\gamma-1}}
\end{equation}
where $\rho_{\mathrm{a}0}$ is the value of the atmosphere density at the spherical radius $R = r_0$. The initial position of the disk surface is 
located whereas the disk and atmosphere pressures are equal.
The initial position of the disk surface is therefore determined by its temperature, defined by $\epsilon$, by the field intensity $\mu$, its
inclination, given by the parameter $m$ and by the density contrast between the disk and the corona $\rho_{\mathrm{a}0}/\rho_0$.

\subsection{Units and normalization}
\label{sec:units}
Since in the formulation of the problem we did not introduce any specific physical scale, the system of Eqs. (\ref{eq:cmass}-\ref{eq:cenergy}) and
the initial conditions presented in Section \ref{sec:incond} can be normalized in arbitrary units: the results will be therefore presented in non-dimensional
units.

Lengths will be given in units of $r_0$, corresponding approximatively to the inner truncation radius of the disk; speeds will be expressed in units of the
Keplerian speed $V_{\mathrm{K}0} = \sqrt{GM/r_0}$ at $r=r_0$ and the densities in units of $\rho_0$, that is the disk initial density at $r=r_0$, $z=0$. 
Assuming for $r_0$ the following units, appropriated for YSO or AGN systems,
\begin{equation}
\begin{array}{lccr}
r_0 & = & 0.1 \; \mathrm{\mbox{AU}} & \qquad \mathrm{(YSO)}\\
      &    &    & \\
      & = & 10\;R_{\mathrm{Schw}} = 10^{-4} \left(\frac{M}{10^8 M_\odot}\right) \; \mathrm{\mbox{pc}} & \qquad \mathrm{(AGN)}
\end{array}
\end{equation}
where $R_{\mathrm{Schw}} = 2GM/c^2$ is the Schwarzschild radius, the Keplerian speed $V_{\mathrm{K}0}$ is given by
\begin{equation}
\begin{array}{lclr}
V_{\mathrm{K}0} & = & 94 \left(\frac{M}{M_\odot}\right)^{1/2} \left( \frac{r_0}{0.1\mathrm{AU}} \right)^{-1/2}\; \mathrm{\mbox{km s}}^{-1} & \qquad \mathrm{(YSO)}\\
                         &   &   & \\
                         & = & 6.7 \times10^4 \left( \frac{r_0}{10 R_\mathrm{Schw}}\right)^{-1/2} \; \mathrm{\mbox{km s}}^{-1}  & \qquad \mathrm{(AGN)}
\end{array}
\end{equation}
Correspondingly time will be expressed in units of $t_0 = r_0/V_{\mathrm{K}0}$:
\begin{equation}
\begin{array}{lccr}
t_0 & = & 1.7 \left(\frac{M}{M_\odot}\right)^{-1/2}  \left( \frac{r_0}{0.1\mathrm{AU}} \right)^{3/2} \; \mathrm{\mbox{days}} & \qquad \mathrm{(YSO)}\\
     &  &  & \\
     & = & 0.5 \left(\frac{M}{10^8 M_\odot}\right) \left( \frac{r_0}{10 R_\mathrm{Schw}}\right)^{3/2} \; \mathrm{\mbox{days}} & \qquad \mathrm{(AGN)}
\end{array}
\end{equation} 
Expressed in units of $t_0$ the Keplerian period of rotation at the inner radius of the disk $r_0$ is equal to $2\pi$.

Finally the normalization density $\rho_0$ can be chosen by determining a suitable mass accretion rate unit $\dot{M}_0 = r_0^2\rho_0V_{\mathrm{K}0}$: 
\begin{equation}
\begin{array}{lcc}
\dot{M}_0 & = & 3\times 10^{-7} \left(\frac{\rho_0}{10^{-12} \; \mathrm{\mbox{g cm}}^{-3}}\right) \left(\frac{M}{M_\odot}\right)^{1/2} \left( \frac{r_0}{0.1\mathrm{AU}} \right)^{3/2} 
                     \; M_\odot \mathrm{\mbox{yr}}^{-1} \\
               &  &   \\
               & = & 9 \left(\frac{\rho_0}{10^{-12} \; \mathrm{\mbox{g cm}}^{-3}}\right) \left(\frac{M}{10^8 M_\odot}\right)^2 \left( \frac{r_0}{10 R_\mathrm{Schw}}\right)^{3/2} 
                      \; M_\odot \mathrm{\mbox{yr}}^{-1} \\
\end{array}  
\label{eq:mrate}
\end{equation}
where, as customary, the first expression corresponds to YSO and the second to AGN objects.
The values of accretion and outflow mass rates will be presented in non-dimensional units and they must be multiplied by the $\dot{M}_0$ factor of Eq. (\ref{eq:mrate}) to 
obtain their physical value.
Similarly torques and powers which will be shown subsequently will be given in units of 
$\dot{J}_0=r_0^3\rho_0V_{\mathrm{K}0}^2$ and $\dot{E}_0=r_0^2\rho_0V_{\mathrm{K}0}^3$ respectively:
\begin{equation}
{\setlength\arraycolsep{1pt}
\begin{array}{lcc}
\dot{J}_0 & = & 3\times 10^{38} \left(\frac{\rho_0}{10^{-12} \; \mathrm{\mbox{g cm}}^{-3}}\right) \left(\frac{M}{M_\odot}\right) \left( \frac{r_0}{0.1\mathrm{AU}} \right)^{2} 
                     \; \mathrm{dyne\; cm} \\
               &  & \\
               & = & 1.2 \times 10^{51}\left(\frac{\rho_0}{10^{-12} \; \mathrm{\mbox{g cm}}^{-3}}\right) \left(\frac{M}{10^8 M_\odot}\right)^3 \left( \frac{r_0}{10 R_\mathrm{Schw}}\right)^{2} 
                      \; \mathrm{dyne\; cm} \\
\end{array}}  
\label{eq:jrate}
\end{equation}

\begin{equation}
{\setlength\arraycolsep{1pt}
\begin{array}{lcc}
\dot{E}_0 & = & 1.9\times 10^{33} \left(\frac{\rho_0}{10^{-12} \; \mathrm{\mbox{g cm}}^{-3}}\right) \left(\frac{M}{M_\odot}\right)^{3/2} \left( \frac{r_0}{0.1\mathrm{AU}} \right)^{1/2} 
                     \; \mathrm{erg\; s}^{-1} \\
               &  & \\
               & = & 2.6 \times 10^{46}\left(\frac{\rho_0}{10^{-12} \; \mathrm{\mbox{g cm}}^{-3}}\right) \left(\frac{M}{10^8 M_\odot}\right)^2 \left( \frac{r_0}{10 R_\mathrm{Schw}}\right)^{1/2} 
                      \; \mathrm{erg\; s}^{-1} \\
\end{array}}  
\label{eq:erate}
\end{equation}
Again, the first equations must be used for YSO while the second for AGN systems.

\subsection{Boundary conditions}
\label{sec:meshbound}
Besides of restricting our study to axisymmetric structures we also assume that the system is symmetric with respect to the midplane of the disk.
The computational domain therefore covers a rectangular region with a radial extent $[0,40r_0]$ and a size along the $z$ direction equal to $[0,120r_0]$: 
axisymmetry is imposed on the rotation axis $r=0$ while planar symmetry is imposed on the disk midplane $z=0$. 
Besides of choosing suitable boundary conditions on the outer sides of the domain, an inner boundary must be placed inside the 
computational box in order to avoid the singularity of the potential well and of the initial setup at the origin of the axis: we therefore define a 
rectangular box with a size $r\times z = [0,r_0]\times[0,0.5r_0]$  which is excluded from the computation and on whose sides boundary conditions 
must be set. 

On the outer right boundary ($r=40r_0$) ``outflow'' conditions, that is zero-gradient, are imposed on thermal pressure, density and poloidal velocity
components. 
On one disk thermal height scale ($z < H r$) the value of the radial velocity in the ghost zones is set to the extrapolated value only if negative and
zero otherwise: in this way we avoid any outflow of matter without imposing the mass accretion rate,  which will  be determined by the dynamical evolution 
of the system.
For $u_\phi$, $B_z$, $B_\phi$ the continuity of the first derivative is required too. The condition on $B_r$ is determined by
imposing the solenoidality of the field, $\nabla\cdot\vec{B}=0$, on the last cell of the domain with a first order approximation.  

On the outer upper boundary ($z=120r_0$) ``outflow'' conditions are imposed on all the variables except for $B_z$ whose boundary values are 
determined to satisfy the $\nabla\cdot\vec{B}=0$ requirement. 

We decided to prescribe the continuity of the first derivative of some variables in the radial direction since the power-law behavior of the initial
conditions along $r$ (see Eq. (\ref{eq:selfsim})) determines strong gradients of the physical quantities on the rightmost boundary. 
This boundary condition affects above all the radial component of the Lorentz force $F_r$:
\begin{equation}
F_r  = -\frac{1}{2}\frac{\partial B_\phi^2}{\partial r}-\frac{B_\phi^2}{r}
                                -\frac{1}{2}\frac{\partial B_z^2}{\partial r}+B_z\frac{\partial B_r}{\partial z}
\label{eq:radlor}
\end{equation}
and therefore the collimation of the outflow.
First of all it is easy to see that a zero-gradient condition on the toroidal magnetic field component $B_\phi$ cancels the pressure gradient of the magnetic
pressure (firs term in the right hand side of Eq. (\ref{eq:radlor})) while the pinching force $-B_\phi^2/r$ (second one) is still present.
An outflow condition on $B_\phi$ on the right outer boundary thus enforces the collimation 
due to the toroidal field. This issue has been widely discussed by Ustyugova et al. (\cite{Ustyugova}) who proposed to use a ``force-free'' condition,
where the poloidal electrical current is parallel to the magnetic field, in order to remove artificial forces originating on the boundaries.
Seen from the point of view of electrical currents, a zero-gradient condition on $B_\phi$ corresponds to have a negative collimating current component 
$J_z$ flowing along the right boundary, while stationary models of accretion-ejection structures show that on the outer part of the disk the current 
outflows from it thus pushing the flow along the field lines without collimating it (see Fig. 13 in Ferreira \cite{Ferreira1997}). 
The continuity of the first derivative of $B_\phi$ on the rightmost boundary allows to develop a gradient of magnetic pressure associated with
$B_\phi$ which can counteract the toroidal pinch and generate an outflowing positive current $J_z$. 

Even if the effects are less pronounced, also the continuity of the first derivative of $B_z$ on the rightmost boundary affects the collimation of the
structure: a pressure gradient directed outwards associated with $B_z$ (third term in the right hand side of Eq. (\ref{eq:radlor}))
counteracts the poloidal field tension (fourth term) thus producing slightly less collimated structures.

On the other hand, we noticed that the choice of an ``outflow'' condition for the toroidal field on the upper boundary did not affect the behavior of the 
outflow as much as the condition on the right boundary.  We did not notice a huge difference between the ``outflow'' condition, where the poloidal current 
has only a $z$ component, and the ``force-free'' condition proposed by Ustyugova et al. (\cite{Ustyugova}), where the poloidal current is parallel to the 
poloidal field.

On the inner boundaries located on the edges of the rectangular region $r\times z = [0,r_0]\times[0,0.5r_0]$, we adopted a similar strategy: 
on the $r = r_0$ side we extrapolated all the physical quantities imposing the continuity of the first derivative except for the poloidal components
of the velocity field, for which a zero-gradient condition was used, and for the radial component of the magnetic field, which is required to fulfill
the $\nabla\cdot\vec{B}=0$ condition; on the $z=0.5r_0$ side we imposed an ``outflow'' condition on all the variables except for the $z$ component
of the field, determined by imposing the solenoidality of the field. 

All the details on the simulation performed are given in the next Section.

\begin{table*}[t]
\caption{Initial parameters of the simulations performed: poloidal resistivity parameter $\alpha_\mathrm{m}$, anisotropy of magnetic resistivity 
$\chi_\mathrm{m}$, cooling parameter $f$, initial accretion rate $\dot{M}/\dot{M}_0$, torque $\dot{J}/\dot{J}_0$ needed to support the accretion rate
between $r_\mathrm{i}=r_0$ and $r_\mathrm{e}=10 r_0$, accretion energy $\dot{E}/\dot{E}_0$ liberated between the same radii, equivalent 
resolution of the adaptive grid.} 
\label{table:cases}
\centering                   
\begin{tabular}{c c c c c c c c}   
\hline\hline        
Simulation & $\alpha_\mathrm{m}$ & $\chi_\mathrm{m}$ & $f$ & $\dot{M}/\dot{M}_0$ & $\dot{J}/\dot{J}_0$ & $\dot{E}/\dot{E}_0$ & Resolution \\   
\hline                  
1 &  0.1 & 1 &  1 & $7.3\times10^{-3}$ & $1.4\times10^{-2}$ & $3.3\times10^{-3}$ & $512\times1536$ \\
2 &  0.1 & 1 &  1 & $7.3\times10^{-3}$ & $1.4\times10^{-2}$ & $3.3\times10^{-3}$ & $128\times384$ \\
3 &  0.1 & 1 &  0 & $7.3\times10^{-3}$ & $1.4\times10^{-2}$ & $3.3\times10^{-3}$ & $512\times1536$ \\
4 &   1   & 1 &  1 & $7.3\times10^{-2}$ & $1.4\times10^{-1}$ & $3.3\times10^{-2}$ & $512\times1536$ \\
5 &   1   & 3 &  1 & $7.3\times10^{-2}$ & $1.4\times10^{-1}$ & $3.3\times10^{-2}$ & $512\times1536$ \\
6 &   1   & 3 &  0 & $7.3\times10^{-2}$ & $1.4\times10^{-1}$ & $3.3\times10^{-2}$ & $512\times1536$ \\
\hline                                   
\end{tabular}
\end{table*}

\subsection{The simulations}
\label{sec:simulations}

Once they are normalized with the units given in Section \ref{sec:units}, the initial conditions presented in Section \ref{sec:incond} depend on 
6 non-dimensional parameters: the ratio between the sound speed and the Keplerian speed at the disk midplane $\epsilon$; the disk
magnetization parameter $\mu$; the magnetic height scale of the initial field $m$; the strength of magnetic diffusivity $\alpha_\mathrm{m}$
and its anisotropy coefficient $\chi_\mathrm{m}$; the ratio between the initial atmosphere and disk densities $\rho_{\mathrm{a}0}/\rho_0$.
All the free parameters except those describing the diffusive properties of the disk will be the same for all the simulations.
We therefore assume a parameter $\epsilon = 0.1$, which fixes the initial thermal height scale and temperature $T$ on the midplane of the disk,
that is:
\begin{equation}
\begin{array}{lcc}
\left. T\right|_{z=0} = \epsilon^2\frac{m_\mathrm{p}GM}{Kr} & = & 10^4 \left(\frac{\epsilon}{0.1}\right)^{2}  \left( \frac{M}{M_\odot} \right) \left(\frac{r}{0.1\mathrm{AU}}\right)^{-1} \; \mathrm{K} \\
                                                                                     & & \\
                                                                                    & = &  5\times 10^{9} \left(\frac{\epsilon}{0.1}\right)^{2} \left(\frac{r}{10 R_\mathrm{Schw}}\right)^{-1} \; \mathrm{K} \\
\end{array}
\end{equation}
The initial structure of the magnetic field is determined by fixing the magnetization parameters $\mu = 0.3$ and $m=0.35$, which gives therefore 
a magnetic height scale $3.5$ times higher than the initial thermal height scale of the disk. We recall that a value of magnetization around equipartition 
or slightly below it is generally required both in numerical (see Zanni et al. \cite{Zanni2004}, Casse \& Keppens \cite{CasKe02}) and analytical (Ferreira \cite{Ferreira1997}, 
Casse \& Ferreira \cite{CasFe00a}) modeling of accretion disks launching jets; this condition is determined by the equilibrium between thermal pressure 
gradients, which vertically support the disk and are responsible for the initial mass loading the field lines, and Lorentz forces, which tend to pinch the disk.
The magnetization parameter $\mu$ fixes the initial value of the poloidal magnetic field on the disk midplane:
\begin{equation}
{\setlength\arraycolsep{1pt}
\begin{array}{cl}
\left. B\right|_{z=0} & = \sqrt{\mu\;8\pi\;P} = \\
 = & 2.6 \left(\frac{\mu}{0.3}\right)^{1/2} \left(\frac{\epsilon}{0.1}\right) \left(\frac{\rho_0}{10^{-12} \; \mathrm{\mbox{g cm}}^{-3}}\right)^{1/2} 
             \left( \frac{M}{M_\odot} \right)^{1/2} \left(\frac{r}{0.1\mathrm{AU}}\right)^{-5/4} \; \mathrm{G} \\
                                               & \\
 = &  1.8\times 10^{3} \left(\frac{\mu}{0.3}\right)^{1/2} \left(\frac{\epsilon}{0.1}\right) \left(\frac{\rho_0}{10^{-12} \; \mathrm{\mbox{g cm}}^{-3}}\right)^{1/2} 
             \left(\frac{r}{10 R_\mathrm{Schw}}\right)^{-5/4} \; \mathrm{G} \\
\end{array}}
\end{equation}
Finally we assumed a density ratio between the corona and the disk $\rho_{\mathrm{a}0}/\rho_0=10^{-4}$.

In order to investigate the effects of magnetic resistivity we performed a series of simulations varying the value of the $\alpha_\mathrm{m}$ parameter
and the anisotropy factor $\chi_ \mathrm{m}$. The summary of all the simulations performed is given in Table \ref{table:cases}.

Isotropic resistive configurations ($\chi_ \mathrm{m} = 1$) have been studied for two different values of the $\alpha_\mathrm{m}$ parameter: 
$\alpha_\mathrm{m}=0.1$ (simulation 1), which corresponds to the magnetic resistivity adopted by Casse \& Keppens (\cite{CasKe02}, \cite{CasKe04}),
and $\alpha_\mathrm{m}=1$ (simulation 4). To determine the effects of an anisotropic resistivity, required by the steady models of Ferreira \& Pelletier (\cite{FerPel1995}),
we performed a simulation characterized by $\alpha_\mathrm{m}=1$ and $\chi_ \mathrm{m} = 3$ (simulation 5): the parameters $\alpha_\mathrm{m}$, 
$\chi_ \mathrm{m}$, $\mu$ and $\epsilon$ of this simulation are typical of cold self-similar solutions found by Casse \& Ferreira (\cite{CasFe00a}).

For these standard simulations the adaptive mesh provided with FLASH is set up with 7 levels of refinement based 
on blocks of $8\times8$ square cells, giving an equivalent resolution of $512\times1536$ points: this resolution is kept fixed in the disk region 
($z< 3\epsilon r$) and around the inner boundary rectangle, while the grid is free to change and adjust the resolution in the outflow region.
In order to determine the importance of numerical diffusive effects in our simulations, we repeated the case characterized by an isotropic $\alpha_\mathrm{m}=0.1$
with a resolution four times lower than the usual one, thus increasing the numerical dissipative effects (simulation 2): for this simulation
the adaptive grid is allowed to reach a maximum equivalent resolution of $128\times384$ points, while maintaining the higher one ($512\times1536$ 
points) around the central inner boundary. The resolution of this test case is similar to the one adopted by Casse \& Keppens (\cite{CasKe02}, \cite{CasKe04}).

This first set of four simulations  is characterized by a cooling factor $f=1$ (Eq. (\ref{eq:cool})) which implies that all the magnetic energy 
Ohmically dissipated is radiated away. 
We therefore repeated simulations 1 and 5 assuming a cooling factor $f=0$ to study the effects of the Ohmic heating in the case of a lower (simulation 3) and
higher resistivity (simulation 6).
The results presented in Section \ref{sec:product}-\ref{sec:statio} will refer to the ``cold''  cases characterized by $f=1$ while a few comments on the ``heated'' 
$f=0$ simulations will be done in Section \ref{sec:joule}.

\begin{figure*}[t]
\centering
\includegraphics[width=17cm]{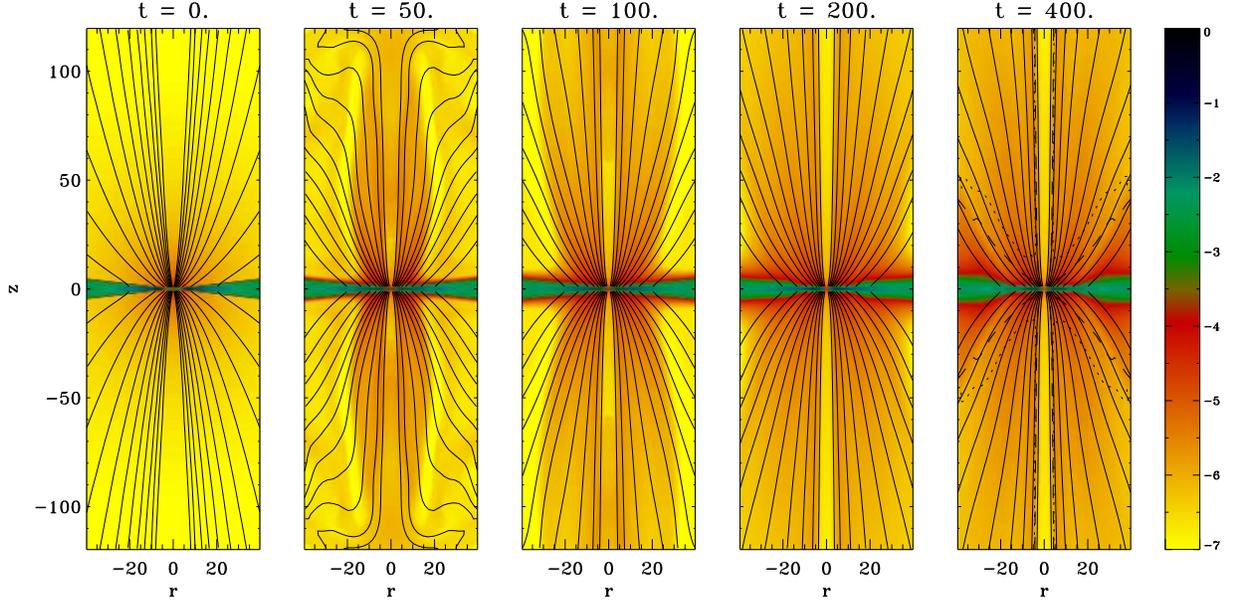}
 \caption{Time evolution of density maps in logarithmic scale of the simulation characterized by ($\alpha_\mathrm{m}=1$, 
$\chi_\mathrm{m}=3$, $f=1$). Time is given in units of $t_0$ (see text). In these units the Keplerian period at the inner radius
of the disk $r=r_0$ is equal to $2\pi$. Superimposed are sample magnetic field lines: the distance between the field lines is proportional
to the intensity of the field. In the last panel ($t=400$) are also plotted the critical Alfven ({\it dashed line}) and fast-magnetosonic 
({\it dotted line}) surfaces. }
 \label{fig:A1T3CT}
\end{figure*}

\begin{figure*}
\centering
\includegraphics[width=17cm]{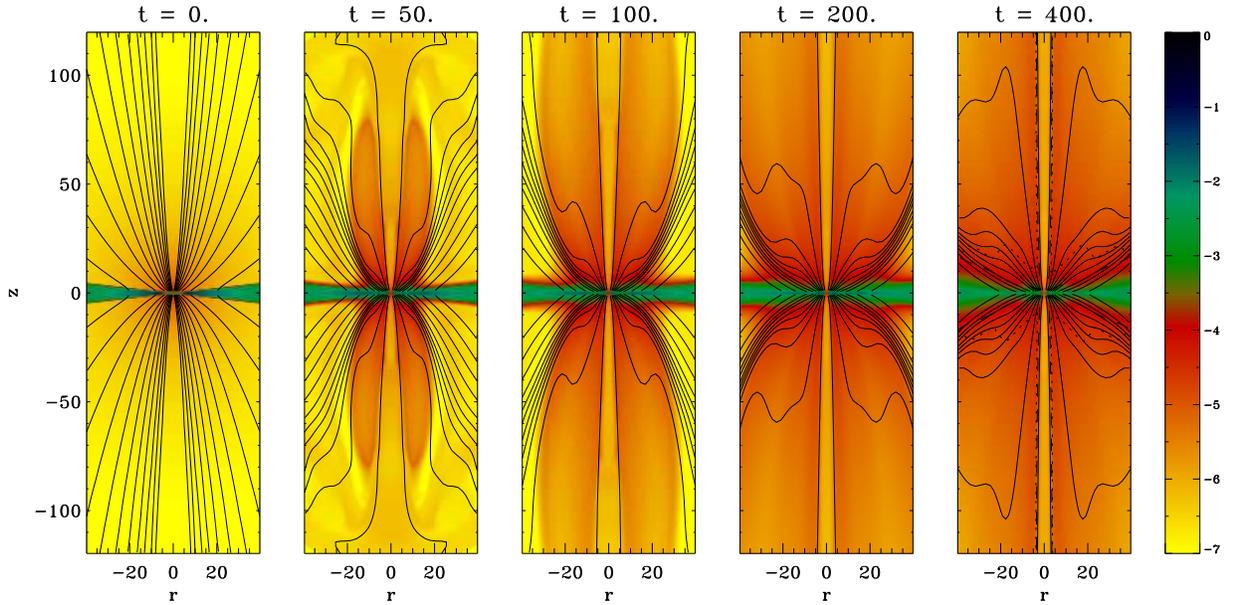}
 \caption{Same as Fig. \ref{fig:A1T3CT} but for the case characterized by  ($\alpha_\mathrm{m}=0.1$, $\chi_\mathrm{m}=1$, $f=1$).}
 \label{fig:A01T1CT}
\end{figure*}

In the fifth column of the Table we show also the mass accretion rate that we imposed at the beginning of our simulations: we recall that this initial rate is determined by
balancing diffusion and advection of the poloidal field inside the disk. Obviously energy and angular momentum must be extracted from the accretion flow to support this
accretion rate: in the sixth and seventh columns of the same Table we show also the torque $\dot{J}$ and the power $\dot{E}$ liberated in the 
accretion between $r_\mathrm{i}=r_0$ and $r_\mathrm{e}=10r_0$. The values of $\dot{J}$ and $\dot{E}$ are with a good approximation given by:
\begin{eqnarray}
\dot{J} & = & \dot{M}\left(\sqrt{GMr_e}-\sqrt{GMr_i}  \right) \\
\label{eq:jdot}
\dot{E} & = & \dot{M}\left(\frac{GM}{2r_i}-\frac{GM}{2r_e} \right)
\label{eq:edot}
\end{eqnarray}
 
\section{Production of jets from magnetized accretion disks}
\label{sec:product}

All the simulations were carried on up to a time $t=400$ which corresponds to $\sim$ 63 periods 
of rotation of the disk at its inner radius. 
On the other hand the final time $t=400$ corresponds to only $0.25$ rotations at the outer boundary $r = 40 r_0$. It
is very unlikely that the outer part of the disk has reached an equilibrium stage at the end of the simulation. Therefore the study of the
accretion-ejection system will be restricted mainly to the inner part of the disk ($r<10r_0$, see Section \ref{sec:mrates}) and to the outflow
coming from it.

In all the cases studied we observed a robust outflow to emerge from the underlying accretion disk:
the solutions show a hollow jet, where the central hole corresponds to the ``sink'' region $r<r_0$.
The outflows are not completely collimated at the end of the runs, that is some matter is outflowing from
the computational box from the outer cylinder at $r=40r_0$. Nevertheless in all the solutions found
the part of the outflow coming from the inner part of the disk crosses the Alfvenic and fast-magnetosonic critical surfaces 
inside the domain (see Fig. \ref{fig:A1T3CT} and \ref{fig:A01T1CT}). Since no disturbance produced in the super-fast region of the outflow can propagate upstream towards the accretion disk,
this condition obviously ensures that the outer boundary conditions do not affect the launching region of this part of the outflow. On the
other hand perturbations can still propagate in a direction transverse to the magnetic field lines: if the fast-Mach cones at the boundaries intersects
the computational box, the boundary conditions can still affect the radial structure and therefore the collimation of the outflow (see Ustyugova et al. 
\cite{Ustyugova}), as already discussed in Section \ref{sec:meshbound}. 

To overcome the problem of the influence of boundary conditions
on the launching phase,  Krasnopolsky et al. (\cite{Krasno2003}) and Anderson et al. (\cite{Anders04}) restricted the launching of the wind to a narrow region
of the inner accretion disk and chose an initial magnetic configuration as to contain the entire fast-magnetosonic critical surface inside the computational domain.
The closed shape of the critical surfaces observed in the cited papers characterizes those solutions more as ``X-winds'' (Shu et al. \cite{Shu1994}), that is 
wide-angle outflows from the corotation radius of a stellar magnetosphere, while the almost conical shape of the critical surfaces of our solutions is typical
of extended disk-winds. Moreover, if on one hand it cannot be taken for granted that an underlying accretion disk can support the magnetic structure and provide 
the mass load imposed as a boundary condition by the authors, on the other hand it has been shown (Ferreira et al. \cite{Ferreira2006}) that ``X-winds'' have 
kinematical properties which are inconsistent with observations of T-Tauri microjets.

\begin{figure}
   \centering
   \includegraphics[width=0.6\linewidth]{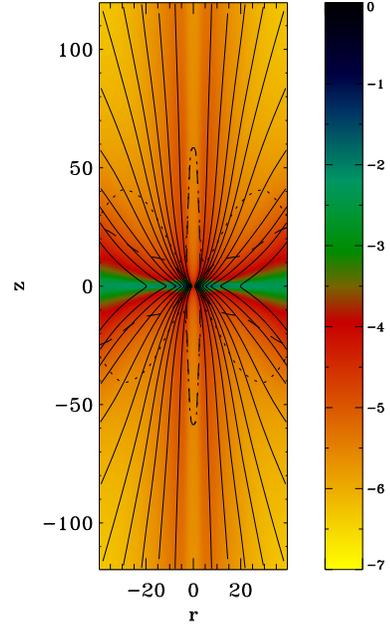}
     \caption{Density map (logarithmic scale) at $t=400$ of the same simulation shown in Fig. \ref{fig:A01T1CT} 
                  performed with a resolution four times smaller.}
         \label{fig:lowres}
 \end{figure}

The structure of the magnetic field allows to distinguish two classes of solutions: the cases characterized by 
$\alpha_\mathrm{m}=1$ show at the end of the computation an ``ordered'' magnetic configuration (Fig. \ref{fig:A1T3CT}),  
while the poloidal field lines are strongly warped and distorted if $\alpha_\mathrm{m}=0.1$ (Fig. \ref{fig:A01T1CT}).
An exception is represented by the case performed with a lower resolution: despite of having a small resistivity parameter 
$\alpha_\mathrm{m}=0.1$, it does not present the characteristic field inversions (Fig. \ref{fig:lowres}). This anomalous behavior can be
reasonably ascribed to the higher numerical dissipation determined by the lower resolution.
 
Other differences can be noticed in Fig. \ref{fig:A1T3CT}-\ref{fig:A01T1CT}: in the more diffusive case the jet is asymptotically less
dense than in the less diffusive one; both outflows become super-Alfvenic ({\it dashed line}) and super-fast-magnetosonic
({\it dotted line}) but in the $\alpha_\mathrm{m}=0.1$ simulation both characteristic surfaces lie closer to the disk.

Despite of these morphological differences between the more and less dissipative cases,  the mechanism which
drives the outflows from the disk is qualitatively the same: in the following Section we will make a few general 
considerations about the forces which determine both the accretion and the ejection flow.

\subsection{Acceleration mechanism}
\label{sec:forces}

Since the accretion-ejection mechanism is mainly magnetically driven it is worth writing explicitly the Lorentz forces
which are acting on the disk-jet system. Following Ferreira (\cite{Ferreira1997}) we decompose the Lorentz force
$\vec{F} = \vec{J} \times \vec{B}$ in the directions parallel and perpendicular to the poloidal field:
\begin{eqnarray}
\nonumber
F_\phi        & = & \frac{B_p}{r}\nabla_\parallel\left(rB_\phi\right) \\
F_\parallel  & = & - \frac{B_\phi}{r}\nabla_\parallel\left(rB_\phi\right) \\
\nonumber
F_\perp      & = & - \frac{B_\phi}{r}\nabla_\perp\left(rB_\phi\right) +J_\phi B_p
\label{eq:lforces}
\end{eqnarray}
where $B_p$ is the poloidal magnetic field while $\nabla_\parallel$ and $\nabla_\perp$ indicate the derivatives parallel and 
perpendicular to the poloidal field respectively. 
These expressions show clearly that the poloidal component of the Lorentz force associated with the toroidal field
is perpendicular to the isosurfaces $rB_\phi$ = const. which are the surfaces along which the poloidal electric current flows.
Moreover they show that the component parallel to the poloidal field $F_\parallel$ and the toroidal one $F_\phi$ are linked by the 
simple relation (see also Casse \& Keppens \cite{CasKe02}, \cite{CasKe04}):
\begin{equation}
F_\parallel = -\frac{B_\phi}{B_p} F_\phi 
\label{eq:torpar}
\end{equation}
(notice that the toroidal field $B_\phi$ assumes negative values in our simulations).
This clearly shows that a poloidal current-field configuration that accelerates the outflow along the field lines ($F_\parallel > 0$) is 
accelerating the plasma also in the toroidal direction ($F_\phi >0$) thus providing an additional centrifugal force. 
This represents the core of the so called magneto-centrifugal mechanism: the magnetic energy stored in the toroidal field at the
base of the outflow both accelerates the plasma along the field lines and increases its angular momentum thus providing a centrifugal
acceleration. As stated by the above relation the relative importance of the two mechanisms, poloidal and centrifugal acceleration, is
given by the ratio $\left|B_\phi\right|/B_p$: when the toroidal field is stronger than the poloidal one, the gradient of $B_\phi$ along the
field lines is the main accelerating mechanism while if $\left|B_\phi\right|/B_p < 1$ the plasma tends to corotate with the mainly poloidal
magnetic field and the centrifugal force is dominant.

 \begin{figure}[t]
   \centering
   \includegraphics[width=\linewidth]{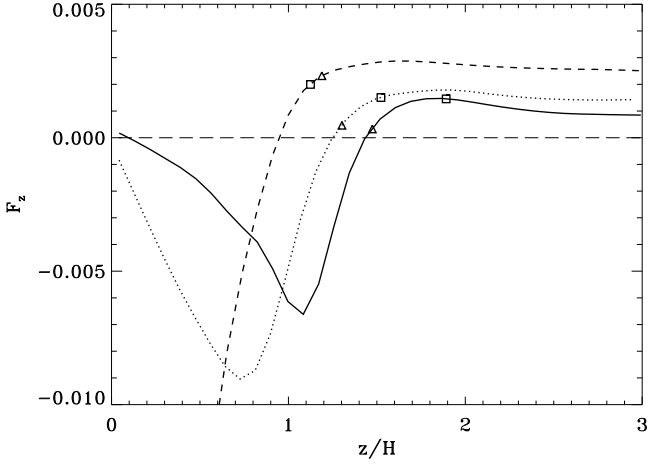}
      \caption{Total force acting on the disk height scale along the $z$ direction for the cases 
                   ($\alpha_\mathrm{m}=1$, $\chi_\mathrm{m}=3$, $f=1$, {\it solid line}),
                   ($\alpha_\mathrm{m}=1$, $\chi_\mathrm{m}=1$, $f=1$, {\it dotted line}) and
                   ($\alpha_\mathrm{m}=0.1$, $\chi_\mathrm{m}=1$, $f=1$, {\it dashed line}). 
                   Along the curves are also indicated the points where the magnetic torque ({\it triangles})
                   and the $z$ component of the Lorentz force({\it squares}) change sign.
                   The curves are calculated at $t=400$.}
         \label{fig:forces}
 \end{figure}
  
Inside the accretion disk the conditions are reversed: as in a unipolar inductor, a radial electric current develops inside the 
conducting plasma, thus providing a toroidal force $-J_r B_z$ which acts to slow down the rotation. According to Eq. (\ref{eq:torpar}), 
since the Lorentz force brakes the matter in the toroidal direction, a poloidal pinching also occurs. 
Therefore, since the disk is vertically pinched by the gravity as well as by the magnetic pressure, only the thermal pressure gradient 
can ensure a quasi-static vertical equilibrium gently lifting the accreting matter towards the surface of the disk where it can be
accelerated to form the outflow. As pointed out in Ferreira (\cite{Ferreira1997}), the transition from accretion to ejection can be therefore 
achieved if the sign of the magnetic torque $F_\phi$ changes sign at the disk surface in order to provide a magnetic acceleration.
  
 \begin{figure}
   \centering
   \includegraphics[width=\linewidth]{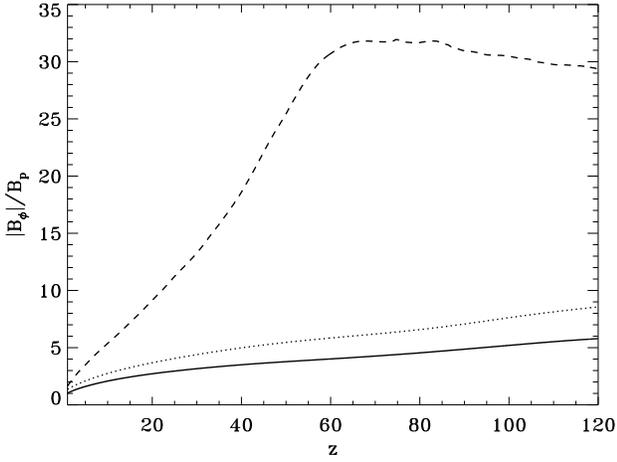}
      \caption{Average ratio between the toroidal and poloidal magnetic fields along the outflows of the simulations characterized by
                   ($\alpha_\mathrm{m}=1$, $\chi_\mathrm{m}=3$, $f=1$, {\it solid line}),
                   ($\alpha_\mathrm{m}=1$, $\chi_\mathrm{m}=1$, $f=1$, {\it dotted line}) and
                   ($\alpha_\mathrm{m}=0.1$, $\chi_\mathrm{m}=1$, $f=1$, {\it dashed line}). The curves are calculated at $t=400$.}
         \label{fig:bratio}
 \end{figure}
 
This scenario is confirmed by our simulations: in Fig. \ref{fig:forces} we plot, at the end of our simulations, the total force, 
given by the sum of gravity, Lorentz force and thermal pressure gradient, acting along the $z$ direction on the disk height scale; 
these curves are obtained by averaging inside the region $r_0<r<10 r_0$ for the cases ($\alpha_\mathrm{m}=1$, $\chi_\mathrm{m}=3$, $f=1$, {\it solid line}), 
($\alpha_\mathrm{m}=1$, $\chi_\mathrm{m}=1$, $f=1$, {\it dotted line}) and ($\alpha_\mathrm{m}=0.1$, $\chi_\mathrm{m}=1$, 
$f=1$, {\it dashed line}). Along each line we also indicated the location of the points where the magnetic torque ({\it triangles}) 
and the total z-component of the Lorentz force ({\it squares}) change their sign.
It is possible to notice that in all the solutions shown the total force changes its sign when the disk is still pinched and braked by the
magnetic field, below both the triangles and the squares: it is therefore the vertical thermal pressure gradient that provides the first
vertical acceleration turning the accretion motion into an outflow; the magnetocentrifugal mechanism becomes effective only in
correspondence of the triangles, where the plasma is accelerated both in the toroidal direction and along the poloidal field lines.

Important differences can be noticed between the three cases shown: decreasing the value of the poloidal or the toroidal magnetic diffusivity 
the location of the points where the total force and the magnetic torque change their sign are located at lower height scales.
A possible explanation of this behavior can be found in Ferreira  (\cite{Ferreira1997}): to change the sign of the magnetic torque it 
is in fact necessary that the radial current $J_r$ responsible for the braking of the disk decreases vertically on a disk scale height;
it has been shown that in a stationary situation the radial current falls off on a height scale proportional to 
$\sqrt{\alpha_\mathrm{m}^2\chi_\mathrm{m}}$ (see Eq. (B1) in Ferreira \cite{Ferreira1997}). Consistently with our simulations, the change
of sign of the torque therefore occurs at higher height scales increasing the value of $\alpha_\mathrm{m}$ or the anisotropy factor
$\chi_\mathrm{m}$. A second noticeable difference is that while in the two higher diffusivity cases ($\alpha_\mathrm{m}=1$) the torque
changes sign when the disk is still pinched by the poloidal magnetic pressure ({\it squares}), in the $\alpha_\mathrm{m}=0.1$ simulation
it changes above this point: in this case the $z$-component of the Lorentz force provides an additional source of mass loading before the 
magnetocentrifugal effect becomes effective. 

Another important difference between these three simulations is shown in Fig. \ref{fig:bratio}, where we plotted a radial average of the 
ratio between the toroidal and poloidal fields $|B_\phi|/B_\mathrm{p}$ along the outflows of the three cases: according to Eq. (\ref{eq:torpar}), 
these curves show that the centrifugal effects are stronger when the magnetic diffusivity is anisotropic ({\it solid line}), while 
for a low diffusivity ({\it dashed line}) the toroidal field pressure gradient is the dominant accelerating force. Moreover the
increase of this ratio with $z$ shows that while near the disk surface the centrifugal force is effectively contributing to the acceleration of the
outflow, it becomes negligible far from it.

\begin{figure*}[t]
\centering
\begin{minipage}[c]{.6\linewidth}
\centering
\includegraphics[width=\linewidth]{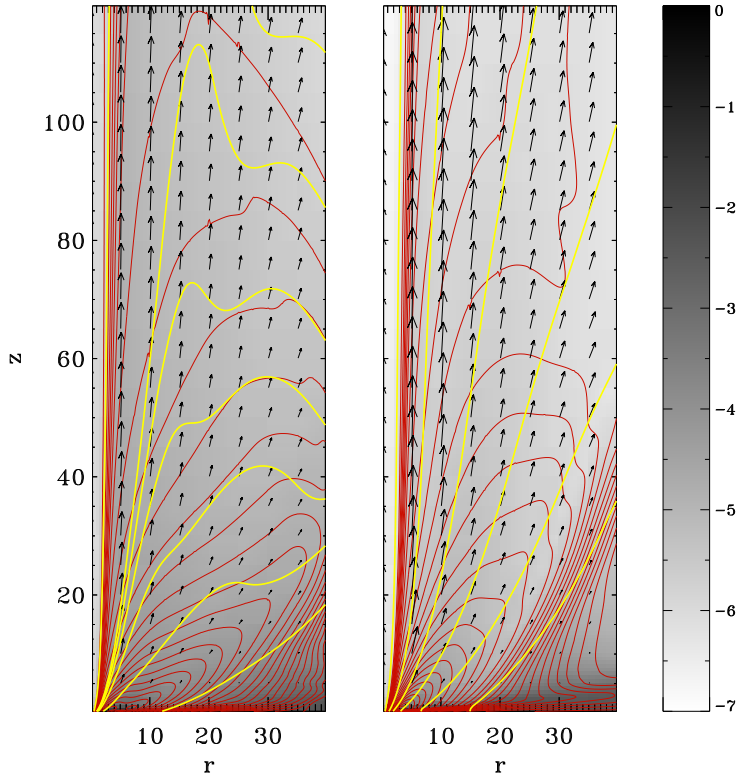}
\end{minipage}
\hfill
\begin{minipage}[c]{.3\linewidth}
\centering
\caption{Poloidal current circuits ({\it solid thin red lines}) at $t=400$ of the cases ($\alpha_\mathrm{m}=0.1$, $\chi_\mathrm{m}=1$, 
            $f=1$, {\it left panel}) and ($\alpha_\mathrm{m}=1$, $\chi_\mathrm{m}=3$, $f=1$, {\it right line}). Plotted
            are also sample poloidal field lines ({\it solid thick yellow lines}) and the poloidal speed vectors. In the background density
            maps in logarithmic grayscale are shown.}
\label{fig:figcurr}
\end{minipage}
\end{figure*}

\subsection{Current circuits}
\label{sec:currents}

\begin{figure*}
\centering
\includegraphics[width=17cm]{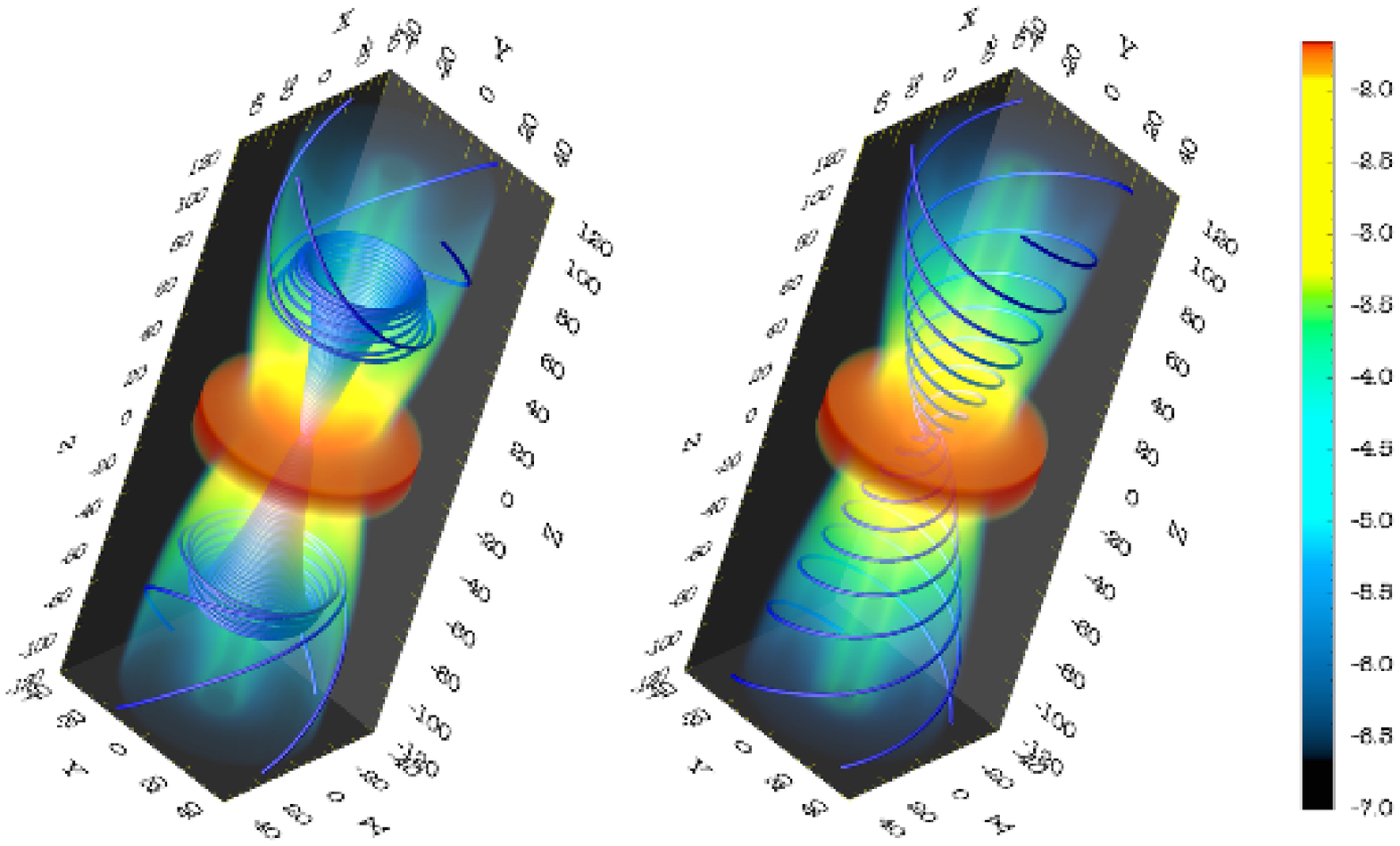}
 \caption{Three dimensional rendering of density at $t=130$ for the cases ($\alpha_\mathrm{m}=0.1$, $\chi_\mathrm{m}=1$, 
            $f=1$, {\it left panel}) and ($\alpha_\mathrm{m}=1$, $\chi_\mathrm{m}=3$, $f=1$, {\it right panel}).
            We also plotted four sample magnetic field lines wrapped around the same magnetic surface.}
 \label{fig:imagebar}
\end{figure*}
A suitable way to understand how the accretion-ejection mechanism works, is to analyze the circulation of the poloidal current 
in the disk-outflow system: in Fig. \ref{fig:figcurr} we show 
the poloidal current circuits ({\it solid thin lines}) superimposed to density maps in logarithmic scale at $t=400$. Sample field lines
({\it solid thick lines}) and the poloidal speed vectors are also plotted; the left panel refers to the ($\alpha_\mathrm{m}=0.1$, 
$\chi_\mathrm{m}=1$, $f=1$) case while the right one to the ($\alpha_\mathrm{m}=1$, $\chi_\mathrm{m}=3$, $f=1$)
simulation. The current circuits are given by the $rB_\phi$ = const. isosurfaces: the poloidal currents circulate counter-clockwise 
and the Lorentz forces are directed outwards perpendicularly to the closed circuits. 
As discussed before, a strong radial positive current flows along the disk midplane, thus providing a braking and vertically pinching force;
at the disk surface the current changes direction until a force component accelerating the plasma along the poloidal field appears.

Some differences can be noticed between the two panels: due to a strong advection determined by the accretion flow, 
the poloidal field lines are much more inclined inside the disk in the less diffusive case (left panel) thus providing a vertically directed
magnetic tension which mass loads the outflow as it was shown in Fig. \ref{fig:forces} (see the position of the square symbol along the dashed line); 
Moreover, it is possible to see that the different shape of the current circuits, together with the higher inclination of the poloidal field lines, 
develops in the $\alpha_\mathrm{m}=0.1$ a strong vertical Lorentz force which strongly bends the field lines, as it was already shown
in Fig. \ref{fig:A01T1CT}. 

This effect is even more evident if we look at a three-dimensional rendering of the two simulations discussed here:
in Fig. \ref{fig:imagebar} we show a 3D rendering of density maps in logarithmic scale for the two cases ($\alpha_\mathrm{m}=0.1$, 
$\chi_\mathrm{m}=1$, $f=1$, {\it left panel}) and ($\alpha_\mathrm{m}=1$, $\chi_\mathrm{m}=3$, $f=1$, {\it right panel}) at a time
$t=130$. It is possible to see that in the more diffusive case the field lines ({\it solid blue lines}) are gently  wrapped around the  magnetic 
surfaces. On the other hand, in the  $\alpha_\mathrm{m}=0.1$ simulation the footpoints of the field lines are advected towards the
central object, not balanced by a strong enough diffusion; a strong differential rotation along the field line therefore appears, 
the footpoint of the line rotating much faster than its opposite end; 
a strong toroidal field develops at the footpoint which gives the strong force directed vertically shown in Fig. \ref{fig:figcurr} that
completely distorts the field lines. 

The mainly toroidal feature visible in the left panel of Fig. \ref{fig:imagebar} propagates vertically along the outflow axis similarly to
a ``magnetic tower'' (Li et al. \cite{Li2001}, Lynden-Bell \cite{Lynbell2003}). Moreover this mechanism is not transient,
repeating itself on every field line when its footpoint is advected towards the faster rotating central part of the disk.
Since the $\alpha_\mathrm{m}=0.1$ simulation at a low resolution shows a magnetic configuration similar to the right panel of
Fig. \ref{fig:imagebar}, it is possible to speculate that the higher numerical diffusion can effectively balance the advection of the 
footpoints so that no strong differential rotation along the field lines is present.
 
Thanks to the current circuits shown in Fig. \ref{fig:figcurr} we can also better understand how the toroidal field can 
act to collimate the outflow against the push of the centrifugal force of the rotating jet. It is in fact usually assumed
that the self-generated toroidal field can automatically ensure the collimation of the outflow through the so called 
``hoop stress'' due to the magnetic tension of the field and equal to $-B_\phi^2/r$. What emerges clearly from both the 
panels of Fig. \ref{fig:figcurr} is that in the outer part of the outflow, where the currents flow out from the disk with a positive
$J_z$ component, the Lorentz force associated with the toroidal field pushes the plasma outwards, thus uncollimating the
jet: in fact, in this situation the outward directed pressure gradient of the toroidal field is stronger than the collimating 
tension. It must be pointed out that the shape of the electric current circuits opposing the collimation of the outer layers 
of the outflow strongly depends on the outer boundary conditions on $B_\phi$, as discussed already in Section 
\ref{sec:meshbound}. On the other hand this behavior is not just a numerical artifact: the current circuits of our 
simulations have the typical butterfly-like shape of analytical models of disk-winds (see Fig. 13 in Ferreira \cite{Ferreira1997}).
The outflow starts to be collimated where the circuit closes back with a negative $J_z$ component so that the ``hoop stress''  exceeds 
the pressure gradient. As the poloidal velocity vectors in Fig. \ref{fig:figcurr} show, in our solutions only the inner part of the outflow is
almost cylindrically collimated inside the computational domain.

\section{Accretion rates and ejection efficiency}
\label{sec:mrates}

\begin{figure}
\resizebox{\hsize}{!}{\includegraphics{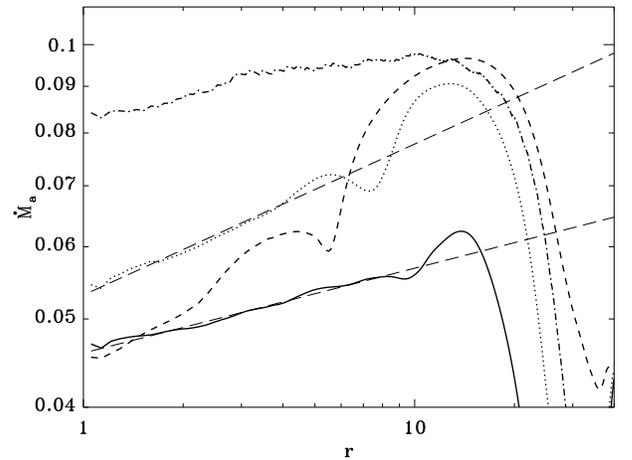}}
\caption{Radial dependence of the accretion rates at $t=400$. The plots refer to the cases with the Ohmic heating suppressed ($f=1$):
            ($\alpha_\mathrm{m}=1$, $\chi_\mathrm{m}=3$ {\it solid line}), ($\alpha_\mathrm{m}=1$, $\chi_\mathrm{m}=1$ {\it dotted line}),
            ($\alpha_\mathrm{m}=0.1$, $\chi_\mathrm{m}=1$ {\it dashed line}) and ($\alpha_\mathrm{m}=0.1$, $\chi_\mathrm{m}=1$ ,
             low resolution {\it dot-dashed line line}). For the two high diffusivity cases a power-law approximation (see text) is also  
            plotted ({\it long-dashed line}).}
\label{fig:macc}
\end{figure}

After having analyzed the acceleration and collimation mechanisms we now take into account the accretion and outflow rates 
which characterize our solutions. In Fig. \ref{fig:macc} we plot for all the four simulations in which we suppressed the Ohmic heating, 
the accretion rates $\dot{M}_\mathrm{a}$ at $t=400$, defined as:
\[
\dot{M}_\mathrm{a} = -2\pi r \int_{-1.6\mathrm{H}}^{1.6\mathrm{H}} \rho u_\mathrm{r} \, \mathrm{d}z
\]
as a function of the cylindrical radius $r$. $\mathrm{H}(r)$ is the thermal height scale of the disk defined as 
$\mathrm{H} = \left. \left(c_\mathrm{s}/\Omega_\mathrm{K}\right)\right|_{z=0}$: the above integral is calculated up to $1.6\mathrm{H}$, 
where the magnetic diffusivity defined by Eq. (\ref{eq:mdiff}) becomes negligible and the plasma, now in an ideal MHD regime, should
flow out from the disk almost parallel to the poloidal magnetic field.
It is possible to see that for all the curves shown, the accretion rate has a sudden decrease for $r\gtrsim 10$: this is due to the
fact that the outflow is extracting efficiently mass and angular momentum only for $r < 10$, which is therefore identified
as the ``launching region''. 
As already pointed out in Section \ref{sec:product} at the end of our simulations the outer part of the disk has still not reached 
a dynamically relevant time scale and therefore had no possibility to fully develop an outflow. 
Moreover the flow emerging from the inner radii of the accretion disk becomes super-fast-magnetosonic inside
the computational domain and therefore the ``launching region'' will not be affected by the outer boundary conditions.
We can therefore define a control volume delimiting this part of the disk: this volume is delimited by an 
inner cylinder of radius $r_\mathrm{i}=1$ and height $-1.6\mathrm{H}(r_\mathrm{i})<z<1.6\mathrm{H}(r_\mathrm{i})$ 
and by an outer cylinder of radius $r_\mathrm{e}=10$ and height $-1.6\mathrm{H}(r_\mathrm{e})<z<1.6\mathrm{H}(r_\mathrm{e})$.
These surfaces will be indicated as  $\vec{S}_\mathrm{i}$ and $\vec{S}_\mathrm{e}$ respectively. Above the disk  the control
volume is closed at the surface determined by the height scale $1.6\mathrm{H}(r)$ between $r_\mathrm{i}=1$ and $r_\mathrm{e}=10$. 
This surface will be indicated as $\vec{S}_\mathrm{s}$.

The four cases shown in Fig. \ref{fig:macc} are characterized by a different outer accretion rate 
(calculated through $\vec{S}_\mathrm{e}$, i.e. at $r_\mathrm{e}=10$): 
when the poloidal resistivity parameter $\alpha_\mathrm{m}$ or the anisotropy coefficient $\chi_\mathrm{m}$ are increased the outer 
accretion rate, indicated as $\dot{M}_\mathrm{ae}$, becomes smaller.  
Moreover it is evident that in the cases characterized by $\alpha_\mathrm{m} = 0.1$ ({\it dotted} and {\it dash-dotted line})
the accretion rate is at least one order of magnitude higher than the initial one (see Table \ref{table:cases}) which was determined to 
balance the advection and the diffusion of poloidal field lines inside the disk: it is therefore clear that in these cases the advection of the
field should dominate thus explaining the peculiar dynamical effects of the $\alpha_\mathrm{m}=0.1$ simulations shown in the previous Section. 
Despite of having such a high accretion rate, we recall that the low resistivity case at low resolution ({\it dash-dotted line}) does not 
show these features, thus supporting the idea that the strong advection is balanced by a high numerical diffusion.

Besides of being characterized by a different $\dot{M}_\mathrm{ae}$, the solutions show also a different slope of the accretion
rate inside the ``launching region''. This slope is clearly linked to the amount of mass that is extracted from the disk and goes into the 
outflow: defining a simplified power-law radial behavior for the accretion rate (Ferreira \& Pelletier \cite{FerPel1995})
\[
\dot{M}_\mathrm{a}(r) = \dot{M}_\mathrm{ae}\left( \frac{r}{r_\mathrm{e}}  \right)^{\xi}
\] 
we can find, imposing mass conservation inside the accretion disk, a simple relation between the ejection efficiency and the ejection
parameter $\xi$:
\begin{equation}
\frac{2\dot{M}_\mathrm{j}}{\dot{M}_\mathrm{ae}} = 1 - \left( \frac{r_\mathrm{i}}{r_\mathrm{e}} \right)^\xi
\label{eq:mrates}
\end{equation}
Here the ejection efficiency $2\dot{M}_\mathrm{j}/\dot{M}_\mathrm{ae}$ is defined as the ratio of the mass outflow from the disk surface
$\vec{S}_\mathrm{s}$ and the outer accretion rate calculated at $r_\mathrm{e} = 10$; the factor 2 takes into account the two bipolar  jets. 
It is easy to see that the higher the ejection efficiency the higher the ejection parameter $\xi$ and therefore the steeper is the slope 
of the accretion rate.
\begin{figure}
\resizebox{\hsize}{!}{\includegraphics{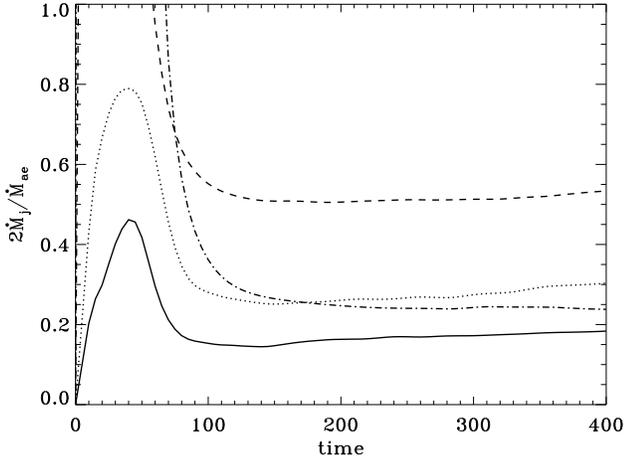}}
\caption{Temporal evolution of the ejection efficiency. The plots refer to the cases
            ($\alpha_\mathrm{m}=1$, $\chi_\mathrm{m}=3$ {\it solid line}), ($\alpha_\mathrm{m}=1$, $\chi_\mathrm{m}=1$ {\it dotted line}),
            ($\alpha_\mathrm{m}=0.1$, $\chi_\mathrm{m}=1$ {\it dashed line}) and ($\alpha_\mathrm{m}=0.1$, $\chi_\mathrm{m}=1$ ,
             low resolution {\it dot-dashed line line}).}
\label{fig:mflux}
\end{figure}

In Fig. \ref{fig:mflux} we plot the ejection efficiency $2\dot{M}_\mathrm{j}/\dot{M}_\mathrm{ae}$ as a function of time for the four simulations
taken into account in this Section: it is possible to see that after an initial transient, this efficiency reaches an almost constant value,
suggesting the attainment of a stationary final state. On the other hand the solutions reach different efficiency values, increasing
from $20\%$ for the ($\alpha_\mathrm{m}=1$, $\chi_\mathrm{m}=3$) case to $55\%$ for the ($\alpha_\mathrm{m}=0.1$, $\chi_\mathrm{m}=1$)
simulation. This result is consistent with the plot shown in Fig. \ref{fig:forces}, where we showed that the forces acting vertically inside the disk
change sign at lower height scales in less diffusive cases: since the disks become denser towards the midplane this determines also a higher
outflow rate. As usual the low resistivity case at low resolution shows an anomalous
behavior ({\it dash-dotted line}) having an ejection efficiency typical of the $\alpha_\mathrm{m}=1$ cases.

Moreover it is clear the direct correlation between the ejection rate and the ratio $|B_\phi|/B_\mathrm{p}$ plotted in Fig. \ref{fig:bratio}.
This means that the main mechanism accelerating the outflow depends strongly on the outflow rate: the less mass loaded jets are efficiently
centrifugally accelerated, while in the more loaded jets the higher inertia  strongly bends the magnetic field in the azimuthal
direction right above the disk surface; as stated by Eq. (\ref{eq:torpar}), the outflow is therefore mainly accelerated by the pressure 
gradients of the toroidal field (see also Anderson et al. \cite{Anders04}).

Consistently the solutions characterized by a higher ejection efficiency show a steeper slope of the accretion rate as shown by Eq. (\ref{eq:mrates}):
as an example, we plotted in Fig. \ref{fig:macc}, the power-laws calculated solving Eq. (\ref{eq:mrates}) for $\xi$ for the cases 
($\alpha_\mathrm{m}=1$, $\chi_\mathrm{m}=3$) and ($\alpha_\mathrm{m}=1$, $\chi_\mathrm{m}=1$) ({\it long-dashed lines}). The ejection
parameters $\xi$ calculated for these two cases are $\xi=0.09$ and $\xi=0.17$ respectively.

\section{Angular momentum transport}
\label{sec:jrates}

\begin{figure*}
   \includegraphics[width=0.3333\textwidth]{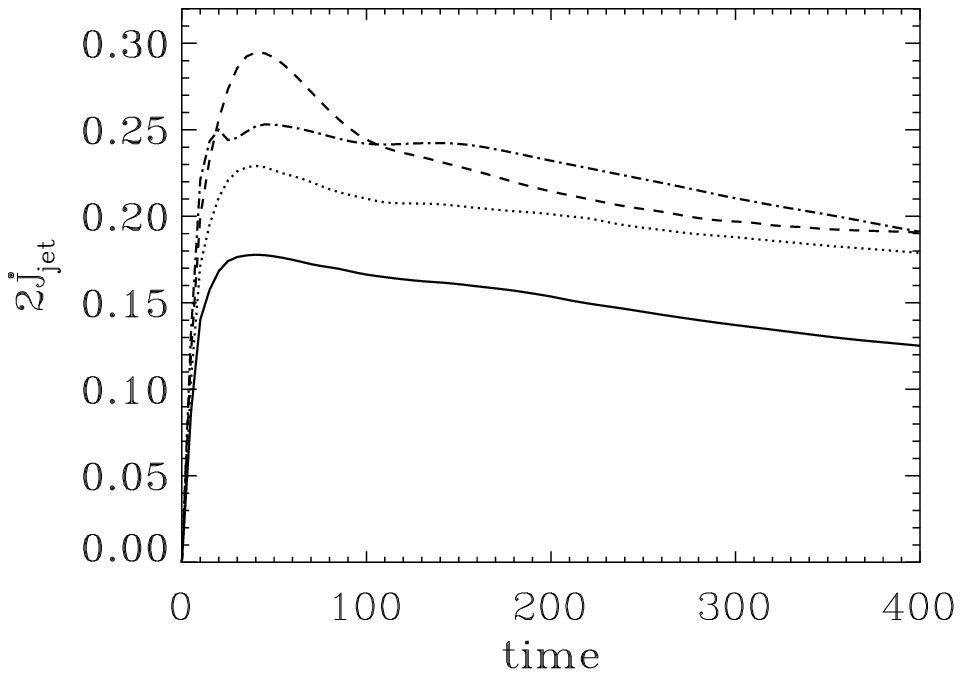}
   \includegraphics[width=0.3333\textwidth]{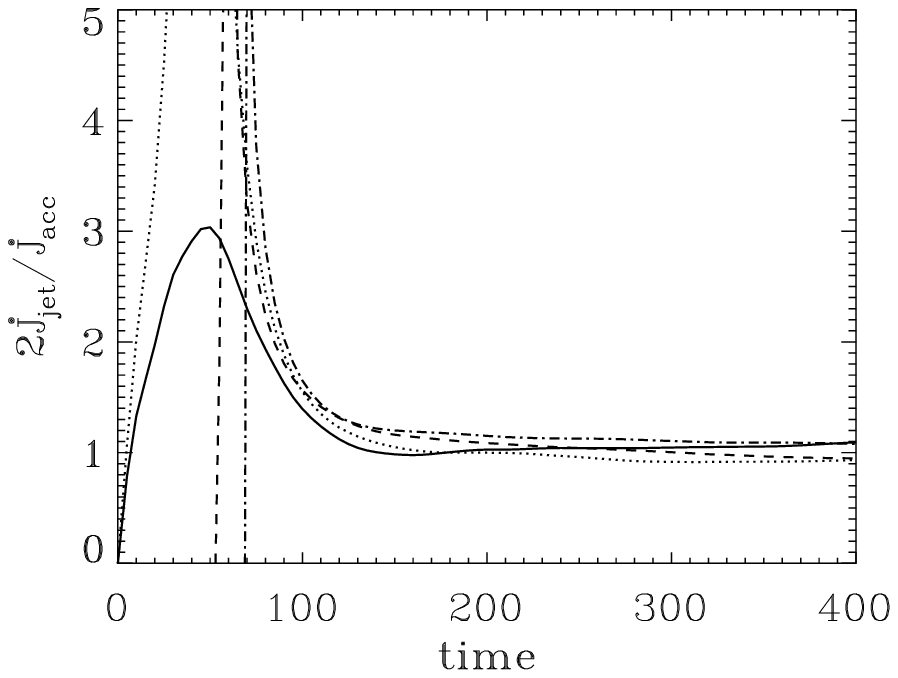}
   \includegraphics[width=0.3333\textwidth]{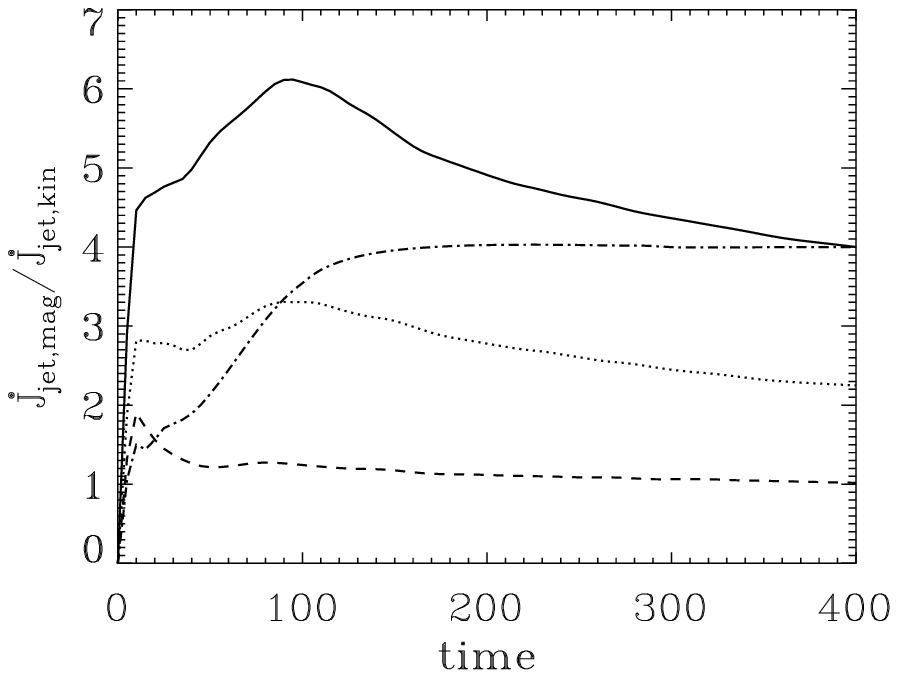}
      \caption{Temporal evolution of the jet torque $2\dot{J}_\mathrm{jet}$ ({\it left panel}), of the ratio between the jet and accretion torque 
                   $2\dot{J}_\mathrm{jet}/\dot{J}_\mathrm{acc}$ ({\it central panel}) and of the ratio between the jet magnetic and mechanical torque
                   $\dot{J}_\mathrm{jet, mag}/\dot{J}_\mathrm{jet, kin}$ ({\it right panel}). The plots refer to the cases
                  ($\alpha_\mathrm{m}=1$, $\chi_\mathrm{m}=3$ {\it solid line}), ($\alpha_\mathrm{m}=1$, $\chi_\mathrm{m}=1$ {\it dotted line}),
                  ($\alpha_\mathrm{m}=0.1$, $\chi_\mathrm{m}=1$ {\it dashed line}) and ($\alpha_\mathrm{m}=0.1$, $\chi_\mathrm{m}=1$ ,
                  low resolution {\it dot-dashed line line}). }
         \label{fig:mombal}
\end{figure*}
It is commonly accepted that disk-winds are a viable mechanism to extract angular momentum from accretion disks without recurring
to viscous torques to allow accretion. In this Section we show how the bipolar jets can extract angular momentum from
the underlying disk and through which channel. 

The angular momentum transport is regulated by the angular momentum flux through the surface delimiting the control volume defined
in Section \ref{sec:mrates}. 
We can therefore define an accretion torque $\dot{J}_\mathrm{acc} = \dot{J}_\mathrm{acc, kin} + \dot{J}_\mathrm{acc, mag}$ defined by the sum
of the two terms:
\begin{equation}
\dot{J}_\mathrm{acc, kin} = \int_{\vec{S}_\mathrm{i}} r_\mathrm{i} \rho u_\phi\vec{u}\cdot\mathrm{d} \vec{S}
                                       -\int_{\vec{S}_\mathrm{e}} r_\mathrm{e} \rho u_\phi\vec{u}\cdot\mathrm{d} \vec{S}
\end{equation}
and
\begin{equation}
\dot{J}_\mathrm{acc, mag} = \int_{\vec{S}_\mathrm{i}} r_\mathrm{i}  B_\phi\vec{B}\cdot\mathrm{d} \vec{S}
                                         -\int_{\vec{S}_\mathrm{e}} r_\mathrm{e}  B_\phi\vec{B}\cdot\mathrm{d} \vec{S}
\end{equation}
For the accretion torque we use a positive sign if it increases the angular momentum contained inside the control
volume and a negative sign if it extracts it.
The term $\dot{J}_\mathrm{acc, kin}$ defines the flux of angular momentum inside the control volume due to the accretion motion,
while $\dot{J}_\mathrm{acc, mag}$ is the magnetic torque which can transport angular momentum 
along the radial direction {\it inside} the disk. This magnetic torque represents a small contribution to $\dot{J}_\mathrm{acc}$:
in all of our simulations the ratio  $\dot{J}_\mathrm{acc, mag}/\dot{J}_\mathrm{acc, kin}$ is always negligible, around -10$\%$.
A small fraction of the disk angular momentum is extracted by the magnetic torque also radially, thus helping 
accretion. On the other hand, the main contribution to $\dot{J}_\mathrm{acc, kin}$ is given by the integral on the outer radius,
where both the accretion rate and the specific angular momentum of the disk are higher. Assuming a Keplerian rotation, a reasonable
approximation for $\dot{J}_\mathrm{acc}$ is therefore given by (see also Eq. (\ref{eq:jdot})):
\begin{equation}
\dot{J}_\mathrm{acc} \sim \dot{M}_\mathrm{ae} \sqrt{GMr_\mathrm{e}}
\label{eq:jdotap}
\end{equation}

In the same way we can define the torque exerted by the outflow on the disk $\dot{J}_\mathrm{jet} = \dot{J}_\mathrm{jet, kin} + 
\dot{J}_\mathrm{jet, mag}$ as the sum of the flux of angular momentum $\dot{J}_\mathrm{jet, kin}$ and of the magnetic torque
$\dot{J}_\mathrm{jet, mag}$  at the disk surface:
\begin{equation}
\dot{J}_\mathrm{jet, kin} = \int_{\vec{S}_\mathrm{s}} r \rho u_\phi\vec{u}\cdot\mathrm{d} \vec{S}
\end{equation}
and
\begin{equation}
\dot{J}_\mathrm{jet, mag} = \int_{\vec{S}_\mathrm{s}} r  B_\phi\vec{B}\cdot\mathrm{d} \vec{S}
\end{equation}
For the jet torque we use a positive sign if extracts angular momentum from the disk.

In a steady situation the accretion and the jet torque should be equal: $2\dot{J}_\mathrm{jet} = \dot{J}_\mathrm{acc}$. Using the 
approximation made in Eq. (\ref{eq:jdotap}), the relation
\begin{equation}
\dot{M}_\mathrm{ae} \sqrt{GMr_\mathrm{e}} \sim 2\dot{J}_\mathrm{jet}
\label{eq:acctor}
\end{equation}
shows clearly that the accretion rate at the outer radius of the launching region $r_\mathrm{e}$ is mainly controlled by the torque
exerted by the outflow on the accretion disk.
In the left panel of Fig. \ref{fig:mombal} we plot the temporal evolution of the total torque $2\dot{J}_\mathrm{jet}$ exerted by the bipolar jets 
of the four simulations performed without Joule heating: it is clear the correlation between the outer accretion rate shown in Fig. \ref{fig:macc}
and the jet torque plotted here, as stated by Eq. (\ref{eq:acctor}).
The high accretion rate of the $\alpha_\mathrm{m}=0.1$ cases, which determines the peculiar behavior of
these solutions, is determined by the high torque exerted by the jet, at least ten times higher than the torque 
needed to maintain the initial rate (see Table \ref{table:cases}).

The curves plotted in the left panel of Fig. \ref{fig:mombal} show a slow decrease in time, thus suggesting that our solutions did
not reach a final steady state. On the other hand, the ratio $2\dot{J}_\mathrm{jet}/\dot{J}_\mathrm{acc}$, plotted in the central 
panel of Fig. \ref{fig:mombal}, stays approximatively constant in time and equal to one, after an initial transient: this shows that
the accretion rate of the disk reacts quickly to the slowly varying jet torque suggesting that our simulations are evolving through 
a series of quasi-stationary states.

Even if the outflows of the different simulations are exerting almost the same torque (left panel, Fig. \ref{fig:mombal}), being a factor two 
smaller just in the case characterized by a high and anisotropic magnetic diffusivity ({\it solid line}), the ratio between the magnetic
$\dot{J}_\mathrm{jet, mag}$ and the mechanical torque $\dot{J}_\mathrm{jet, kin}$ shows noticeable differences (right panel in Fig.
\ref{fig:mombal}).
The ratio goes from a value around unity for the less dissipative case ($\alpha_\mathrm{m}=0.1$, $\chi_\mathrm{m}=1$)  up to a 
value $\gtrsim 4$ for the ($\alpha_\mathrm{m}=1$, $\chi_\mathrm{m}=3$) simulation. Once again we point out that the low resistivity
simulation at low resolution ({\it dot-dashed line}) behaves like a higher resistivity case, 
showing $\dot{J}_\mathrm{jet, mag}/\dot{J}_\mathrm{jet, kin}\sim 4$.
This quantity has an easy interpretation:
the angular momentum extracted from the disk is in fact stored initially in the toroidal magnetic field and it is then transferred
starting from the disk surface to the outflowing plasma which is therefore centrifugally accelerated. The ratio
$\dot{J}_\mathrm{jet, mag}/\dot{J}_\mathrm{jet, kin}$ is therefore a measure of the efficiency of the magneto-centrifugal
mechanism: the higher is this ratio, the more specific angular momentum is available at the disk surface, centrifugally 
accelerating the plasma to higher poloidal terminal speeds (see Section \ref{sec:enerbud}).

\section{Energy budget of the disk-jet system}
\label{sec:enerbud}

 \begin{figure*}
   \includegraphics[width=0.3333\textwidth]{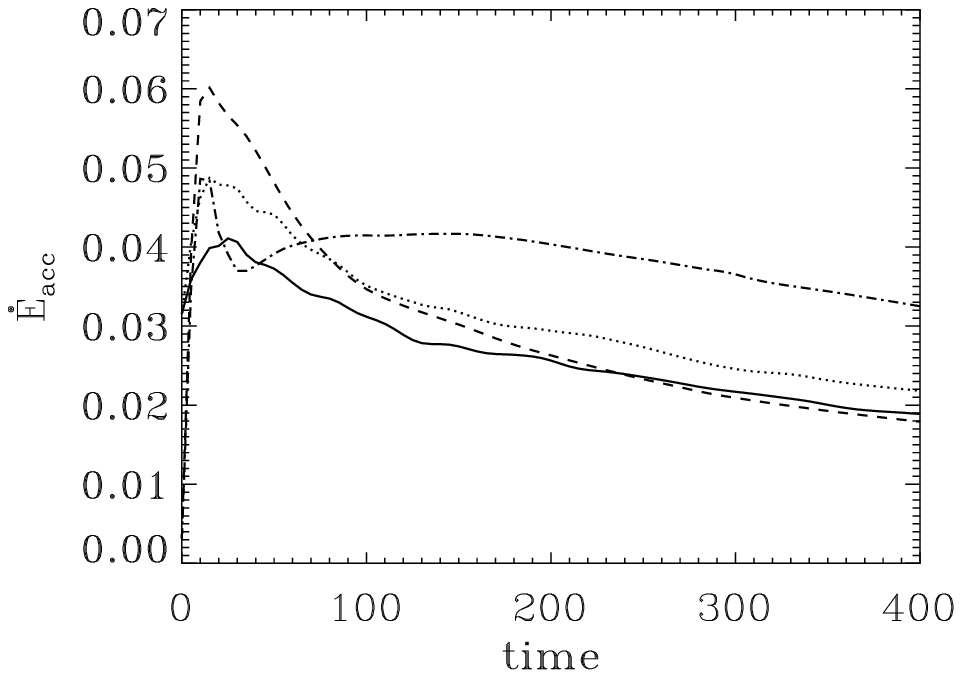}
   \includegraphics[width=0.3333\textwidth]{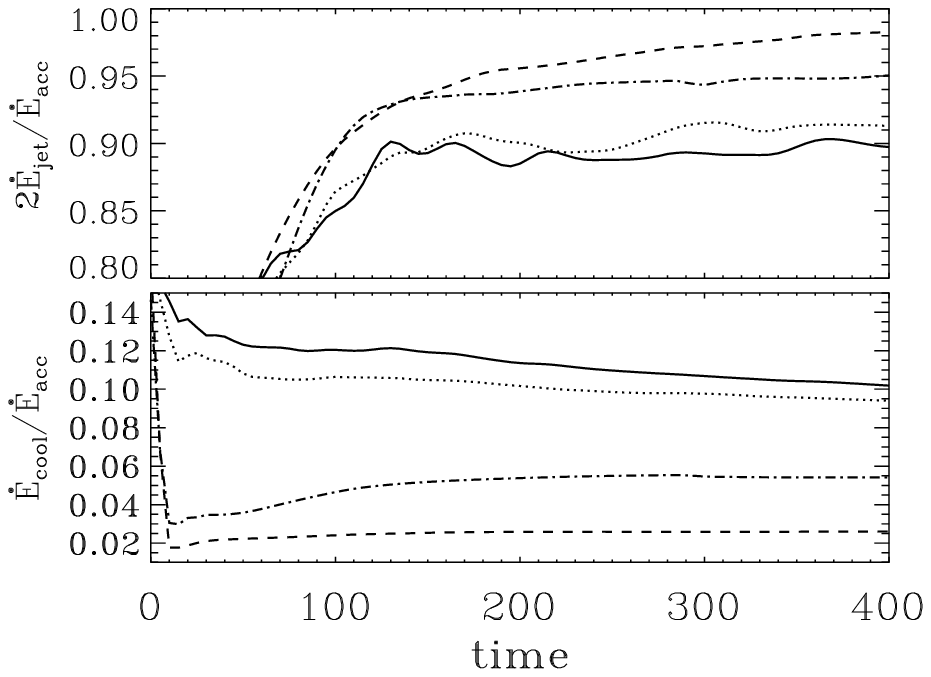}
   \includegraphics[width=0.3333\textwidth]{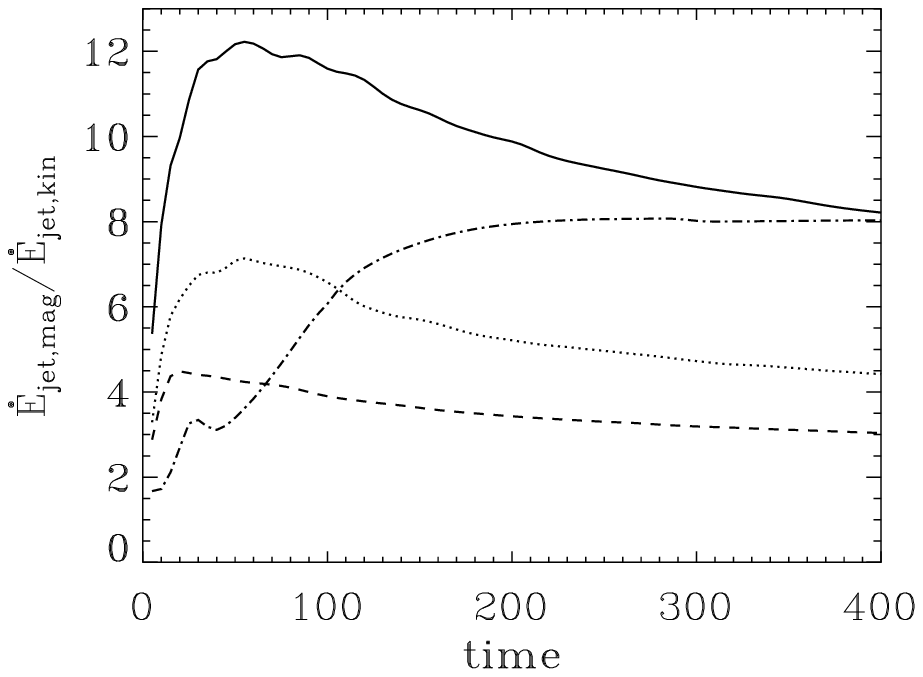}
      \caption{Temporal evolution of the accretion power  $\dot{E}_\mathrm{acc}$ ({\it left panel}); the ratio between the jet and accretion power
                   $2\dot{E}_\mathrm{jet}/\dot{E}_\mathrm{acc}$ and the ratio between the radiated and accretion power
                   $2\dot{E}_\mathrm{jet}/\dot{E}_\mathrm{acc}$ ({\it central panel}); the ratio between the jet magnetic and kinetic power
                   $\dot{E}_\mathrm{jet, mag}/\dot{E}_\mathrm{jet, kin}$ ({\it right panel}). The plots refer to the cases
                  ($\alpha_\mathrm{m}=1$, $\chi_\mathrm{m}=3$ {\it solid line}), ($\alpha_\mathrm{m}=1$, $\chi_\mathrm{m}=1$ {\it dotted line}),
                  ($\alpha_\mathrm{m}=0.1$, $\chi_\mathrm{m}=1$ {\it dashed line}) and ($\alpha_\mathrm{m}=0.1$, $\chi_\mathrm{m}=1$ ,
                  low resolution {\it dot-dashed line line}). }
         \label{fig:enerbal}
 \end{figure*}
In this Section we finally study the energetics of the disk-jet system.
In order to study the energy balance of our simulations we consider the energy fluxes through the surfaces of the control volume
defined in Section \ref{sec:mrates}. We then define an accretion power 
$\dot{E}_\mathrm{acc} = \dot{E}_\mathrm{acc, mec} +  \dot{E}_\mathrm{acc, mag} + \dot{E}_\mathrm{acc, thm}$ given by the
sum of the terms:
\begin{eqnarray}
\dot{E}_\mathrm{acc, mec} & = & \int_{\vec{S}_\mathrm{i}} \left(\frac{u^2}{2} +\Phi_\mathrm{g}\right) \rho \vec{u}\cdot\mathrm{d} \vec{S}
                                                -\int_{\vec{S}_\mathrm{e}}  \left(\frac{u^2}{2} +\Phi_\mathrm{g}\right) \rho \vec{u}\cdot\mathrm{d} \vec{S} \\
\dot{E}_\mathrm{acc, mag} & = & \int_{\vec{S}_\mathrm{i}} \vec{E} \times \vec{B}\cdot\mathrm{d} \vec{S}
                                               - \int_{\vec{S}_\mathrm{e}} \vec{E} \times \vec{B}\cdot\mathrm{d} \vec{S}  \\
\dot{E}_\mathrm{acc, thm} & = & \int_{\vec{S}_\mathrm{i}}  \frac{\gamma}{\gamma-1} P \vec{u}\cdot\mathrm{d} \vec{S}                                               
                                                -\int_{\vec{S}_\mathrm{e}}  \frac{\gamma}{\gamma-1} P \vec{u}\cdot\mathrm{d} \vec{S}                                                                                                    
\end{eqnarray}
where $\dot{E}_\mathrm{acc, mec},  \dot{E}_\mathrm{acc, mag}$ and $\dot{E}_\mathrm{acc, thm}$ represent the mechanical
(kinetic $+$ gravitational), Poynting and enthalpy flux through the cylindrical surfaces at the external and internal radius
of the launching region. For the accretion power we will use a positive sign if it increases the energy contained inside the control volume.
\begin{table}
\caption{Values of the different contributions to the total accretion power $\dot{E}_\mathrm{acc}$ calculated at $t=400$. The simulations
             are characterized by their diffusivity parameters $\alpha_\mathrm{m}$ and $\chi_\mathrm{m})$.  } 
\label{table:enacc}
\centering                   
\begin{tabular}{c@{}l c c c}   
\hline\hline        
Simulation  & & 
\raisebox{0pt}[0pt][0pt]{\raisebox{-2mm}{$\dot{E}_\mathrm{acc, mec}/\dot{E}_\mathrm{acc}$}} & 
\raisebox{0pt}[0pt][0pt]{\raisebox{-2mm}{$\dot{E}_\mathrm{acc, mag}/\dot{E}_\mathrm{acc}$}} & 
\raisebox{0pt}[0pt][0pt]{\raisebox{-2mm}{$\dot{E}_\mathrm{acc, thm}/\dot{E}_\mathrm{acc}$}} \\    
($\alpha_\mathrm{m}$,$\chi_\mathrm{m})$ & & & \\
\hline                  
(0.1,1) &                                           &  0.97 & 0.14  &  -0.11 \\
(0.1,1) & \hspace{-2mm}{\tiny low r.}   &  1.29 & 0.09  &  -0.38 \\
(1,1)    &                                           &  0.97 & 0.10  &  -0.07 \\
(1,3)    &                                           &  0.99 & 0.05  &  -0.04 \\
\hline                                   
\end{tabular}
\end{table}
\begin{table*}
\caption{Values of the different contributions to the total jet power $\dot{E}_\mathrm{jet}$ calculated at $t=400$. The simulations
             are characterized by their diffusivity parameters $\alpha_\mathrm{m}$ and $\chi_\mathrm{m})$. } 
\label{table:ejet}
\centering                   
\begin{tabular}{c@{}l c c c c}   
\hline\hline        
Simulation  & & 
\raisebox{0pt}[0pt][0pt]{\raisebox{-2mm}{$\dot{E}_\mathrm{jet, kin}/\dot{E}_\mathrm{jet}$}} & 
\raisebox{0pt}[0pt][0pt]{\raisebox{-2mm}{$\dot{E}_\mathrm{jet, grv}/\dot{E}_\mathrm{jet}$}} &
\raisebox{0pt}[0pt][0pt]{\raisebox{-2mm}{$\dot{E}_\mathrm{jet, mag}/\dot{E}_\mathrm{jet}$}} & 
\raisebox{0pt}[0pt][0pt]{\raisebox{-2mm}{$\dot{E}_\mathrm{jet, thm}/\dot{E}_\mathrm{jet}$}} \\    
($\alpha_\mathrm{m}$,$\chi_\mathrm{m})$ & & & \\
\hline                  
(0.1,1) &                                           &  0.42 & -0.73  &  1.28 & 0.03\\
(0.1,1) & \hspace{-2mm}{\tiny low r.}   &  0.13 & -0.24  &  1.05 & 0.06\\
(1,1)    &                                           &  0.25 & -0.49  &  1.22 & 0.02\\
(1,3)    &                                           &  0.13 & -0.26  &  1.12 & 0.01 \\
\hline                                   
\end{tabular}
\end{table*}

In the left panel of Fig. \ref{fig:enerbal} we plot the temporal evolution of the total accretion power while in Table 
\ref{table:enacc} we show the different contribution to $\dot{E}_\mathrm{acc}$ at $t=400$ for the cases taken
into account in this Section. We can see that the dominant term is always the mechanical power liberated in
the accretion: the Poynting flux increases slightly the total energy contained in the control volume, while the
enthalpy flux always acts to decrease the accretion power, advecting the thermal energy at the inner radius.
It is important to notice that the simulation with a low magnetic diffusivity performed at a lower resolution (second
line) shows a higher enthalpy flux due to the higher numerical dissipation.
Neglecting the enthalpy and the magnetic contribution and assuming a Keplerian rotation, the energy liberated
by accretion can be safely approximated by (see also Eq. (\ref{eq:edot})):
\begin{equation}
\dot{E}_\mathrm{acc} \sim \dot{M}_\mathrm{ai}\frac{GM}{2r_\mathrm{i}} - 
                                      \dot{M}_\mathrm{ae}\frac{GM}{2r_\mathrm{e}} 
\end{equation}
This approximation shows clearly that the liberated power is mainly determined by the inner accretion rate: the
values of $\dot{E}_\mathrm{acc}$ of our simulations at $t=400$ (left panel of Fig. \ref{fig:enerbal}) are clearly
correlated with the inner accretion rates visible in Fig. \ref{fig:macc}.

To characterize the energy extracted by the outflow we define a jet power $\dot{E}_\mathrm{jet} = 
\dot{E}_\mathrm{jet, kin} + \dot{E}_\mathrm{jet, grv} + \dot{E}_\mathrm{jet, mag} + \dot{E}_\mathrm{jet, thm}$
given by the sum of the fluxes of kinetic, gravitational, magnetic and thermal energy at the disk surface $\vec{S}_\mathrm{s}$:
\begin{eqnarray}
\dot{E}_\mathrm{jet, kin} & = &  \int_{\vec{S}_\mathrm{s}} \frac{u^2}{2} \rho \vec{u}\cdot\mathrm{d} \vec{S} \\
\dot{E}_\mathrm{jet, grv} & = &  \int_{\vec{S}_\mathrm{s}} \Phi_\mathrm{g} \rho \vec{u}\cdot\mathrm{d} \vec{S} \\
\dot{E}_\mathrm{jet, mag} & = &  \int_{\vec{S}_\mathrm{s}}  \vec{E} \times \vec{B} \cdot\mathrm{d} \vec{S} \\
\dot{E}_\mathrm{jet, thm} & = &  \int_{\vec{S}_\mathrm{s}}  \frac{\gamma}{\gamma-1} P  \vec{u}\cdot\mathrm{d} \vec{S} 
\end{eqnarray}
For the jet power we will use a positive sign if it extracts energy from the control volume.

Since in this Section we are taking into account cases characterized by $f=1$, we can also define the radiated power  $\dot{E}_\mathrm{cool}$ as
\begin{equation}
\dot{E}_\mathrm{cool} = \int_{V_\mathrm{c}} \Lambda_\mathrm{cool} \mathrm{dV} =  \int_{V_\mathrm{c}} \bar{\bar{\vec{\eta}}}\vec{J}\cdot\vec{J} \mathrm{dV}
\end{equation}
where $V_\mathrm{c}$ is the control volume.
In a steady situation the energy fluxes should balance to give $2\dot{E}_\mathrm{jet} = \dot{E}_\mathrm{acc} - \dot{E}_\mathrm{cool}$. 
In the central panel of Fig. \ref{fig:enerbal} we plot the time evolution of the ratios $2\dot{E}_\mathrm{jet}/\dot{E}_\mathrm{acc}$ and
$\dot{E}_\mathrm{cool}/\dot{E}_\mathrm{acc}$. We can see that most of the accretion power is liberated in the jet,
increasing from $\sim90\%$ for the more dissipative case ($\alpha_\mathrm{m}=1$, $\chi_\mathrm{m}=3$) up to $\sim 98\%$ for the
less dissipative one ($\alpha_\mathrm{m}=0.1$, $\chi_\mathrm{m}=1$). Correspondingly the radiated power decreases from $\sim 10\%$
down to $\sim2\%$ respectively. Anyway the two contibutions sum up to approximatively one, suggesting a quasi-stationary situation.

As it was noticed for the jet torques, also the jet powers do not show a huge difference between the simulations. On the other hand
the different contributions to the total jet power depend more clearly on the simulation parameters. In Table \ref{table:ejet} we see
that the greater contribution to the jet power at the surface of the disk comes from the Poynting flux, since most of the energy is
stored in the toroidal magnetic field. The enthalpy flux is always negligible: our jets are cold, undergoing a continuous adiabatic 
expansion starting from the disk surface.
The kinetic and the gravitational fluxes are strongly correlated with the outflow rates of the different cases: the gravitational 
contribution is negative since at the disk surface the outflow is still inside the potential well of the central object.

\begin{figure}
\resizebox{\hsize}{!}{\includegraphics{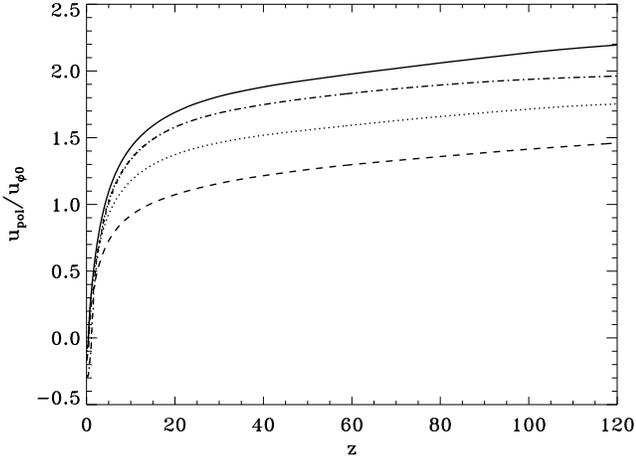}}
\caption{Average poloidal speed $\mathrm{u}_\mathrm{pol}$ along the jets calculated at $t=400$ normalized to the value of 
            the toroidal (Keplerian) speed $\mathrm{u}_{\phi0}$ at the base of each streamline. These curves have been obtained 
            averaging the ratio $\mathrm{u}_\mathrm{pol}/\mathrm{u}_{\phi0}$ on all the streamlines outflowing from the upper 
            boundary of the computational box. The line style used refers to the same cases as in Fig. \ref{fig:enerbal}.}
\label{fig:velz}
\end{figure}

In the right panel of Fig. \ref{fig:enerbal} we plot the temporal evolution of the ratio between the jet Poynting and kinetic flux
$\dot{E}_\mathrm{jet, mag}/\dot{E}_\mathrm{jet, kin}$. 
A higher $\dot{E}_\mathrm{jet, mag}/\dot{E}_\mathrm{jet, kin}$ ratio indicates that more magnetic energy per particle is available
at the disk surface to accelerate the outflow. 
The Poynting flux available at the disk surface is then converted into kinetic energy along the outflow and at the upper end of
the computational box the kinetic flux is dominant, as confirmed by the fact that the flow is super-fast-magnetosonic. The asymptotic speed of the
outflow is therefore higher for cases characterized by a higher $\dot{E}_\mathrm{jet, mag}/\dot{E}_\mathrm{jet, kin}$, as confirmed  by Fig. \ref{fig:velz} 
where we plot an average value of the poloidal speed normalized to the toroidal (Keplerian) speed at the base of each streamline. 
The poloidal velocity measured at the upper end of the computational domain is a few times the Keplerian speed at the base of the streamlines, 
consistently with the fact that different classes of astrophysical jets show velocities of the same order of the escape velocity from the central object (Livio \cite{Livio}).

\begin{figure}
\resizebox{\hsize}{!}{\includegraphics{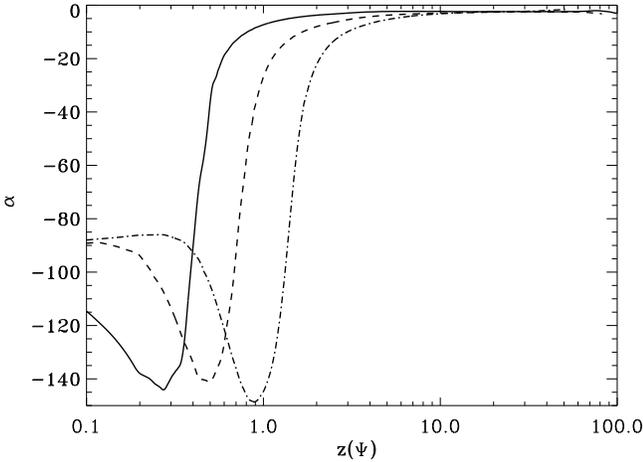}}
\caption{Angle between poloidal speed and poloidal magnetic field for the case characterized by $\alpha_\mathrm{m}=1$, 
             $\chi_\mathrm{m}=3$ and with the Ohmic heating suppressed. This quantity is calculated at $t=400$ along three 
             sample field lines whose footpoints are located at  $r_0$ = 2 ({\it solid line}), 4 ({\it dashed line}) 
             and 8 ({\it dot-dashed line}).}
\label{fig:angle}
\end{figure}

\section{A quasi-steady solution}
\label{sec:statio}

In this Section we will try to characterize one of the simulations performed as a steady solution. We already pointed out that none of
our solutions are perfectly steady: the low diffusivity case shows a continuously evolving magnetic structure,  while
all the simulations show a slowly evolving jet torque (left panel in Fig. \ref{fig:mombal}) and accretion power (left panel in Fig. 
\ref{fig:enerbal}). On the other hand the constant values assumed by the accretion efficiency (Fig. \ref{fig:mflux}), by the
ratio between the jet and the accretion torque (central panel in Fig. \ref{fig:mombal}) and by the ratio between the 
jet and the accretion power   (central panel in Fig. \ref{fig:enerbal}), indicate that our solutions, at least those with 
$\alpha_\mathrm{m}=1$, are slowly evolving through a series of quasi-stationary states. In this Section we will analyze the 
simulation characterized by $\alpha_\mathrm{m}=1$, $\chi_\mathrm{m}=3$ and $f=1$: besides of 
showing some features of a stationary solution, its parameters $\alpha_\mathrm{m}$, $\chi_\mathrm{m}$, $\mu$ and $\epsilon$ 
are typical of the cold self-similar steady solutions found by Casse \& Ferreira (\cite{CasFe00a}), allowing a direct comparison between steady
and time-dependent solutions.

We recall a few relations which are valid for a steady axisymmetric solution of the ideal MHD equations. The poloidal speed 
$\vec{u}_\mathrm{p}$ is related to the poloidal magnetic field $\vec{B}_\mathrm{p}$ 
\begin{equation}
\vec{u}_\mathrm{p} = \frac{K\left(\Psi\right)}{\rho}\vec{B}_\mathrm{p} 
\end{equation}
where $K\left(\Psi\right)$ is a constant along every field line marked by the flux function $\Psi$ and it represents the ratio between 
the mass flux and the magnetic flux. This relation clearly states that in a steady situation the poloidal speed and magnetic field are 
parallel. The toroidal speed $u_\phi$ is related to the rotation rate of the magnetic surface $\Omega(\Psi)$ by
\begin{equation}
u_\phi = \Omega(\Psi)r+\frac{K\left(\Psi\right)}{\rho}B_\phi
\end{equation}
The total specific angular momentum $l(\Psi)$, which is constant along each field line, is given by
\begin{equation}
l(\Psi) = ru_\phi-\frac{rB_\phi}{K(\Psi)} = \Omega r_\mathrm{A}^2
\end{equation}
where $r_\mathrm{A}$ is the Alfven radius. Each field line is therefore characterized by the non-dimensional constants $k$, or 
mass loading parameter, and $\lambda$, which gives a measure of the magnetic lever arm $r_\mathrm{A}/r_0$:
\begin{eqnarray}
k & = & K(\Psi)\frac{r_0\Omega}{B_0} \\
\lambda & = & \frac{l(\Psi)}{\Omega r_0} = \left(\frac{r_\mathrm{A}}{r_0}\right)^2
\end{eqnarray}
where $r_0$ is the cylindrical radius of the footpoint of the field line and $B_0$ is the value of the magnetic field at $r_0$.

\begin{figure}
\resizebox{\hsize}{!}{\includegraphics{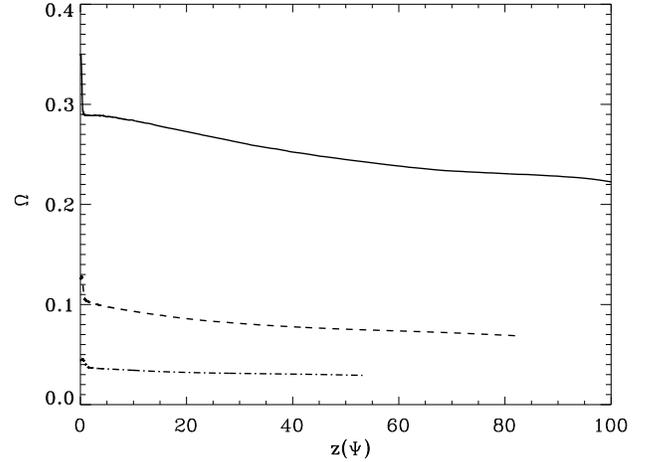}}
\caption{Rotation rate of the magnetic field calculated along the fieldlines whose footpoints are located at  $r_0$ = 2  ({\it solid line}), 
             4 ({\it dashed line}) and 8 ({\it dot-dashed line}.}
\label{fig:omega}
\end{figure}

We therefore tried to characterize our simulation computing these steady invariants along three sample
field lines whose footpoints are located at $r_0$ = 2 ({\it solid line} in the next Figures), 4 ({\it dashed line}) and 8 
({\it dot-dashed line}).
In Fig. \ref{fig:angle} is plotted the angle between the poloidal field and the poloidal speed: while inside the disk, where the
magnetic resistivity is effective, the accreting plasma flows perpendicularly to the field lines, in the outflow, where the
ideal MHD holds, the matter flows along the lines.
In Fig. \ref{fig:omega} is plotted the $\Omega$ invariant: the rotation rate of the field lines, defined by a reference frame in
which the electric field is zero, is close to the Keplerian rate, since the field lines are anchored in the accretion disk and
corotate with it. Moreover it is possible to see that $\Omega$ is slightly higher at the base of the outflow: as in the low
resistivity case but without noticeable consequences, the footpoint of the field line is advected towards the central part of the
disk, rotating a little bit faster than its outer end.

\begin{figure}
\resizebox{\hsize}{!}{\includegraphics{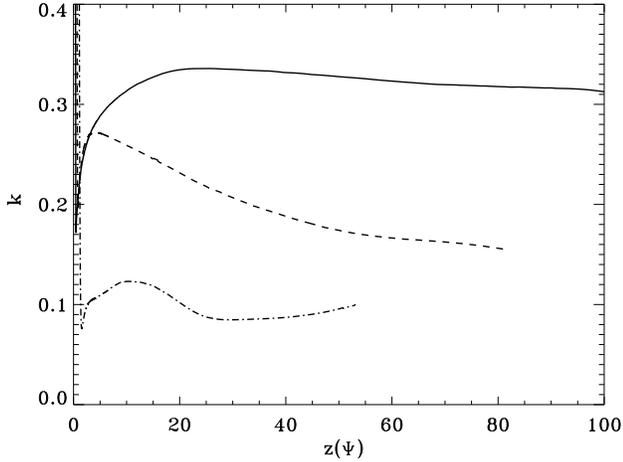}}
\caption{Mass loading parameter calculated along the fieldlines whose footpoints are located at  $r_0$ = 2  ({\it solid line}), 
             4 ({\it dashed line}) and 8 ({\it dot-dashed line}).}
\label{fig:kappa}
\end{figure}

\begin{figure}
\resizebox{\hsize}{!}{\includegraphics{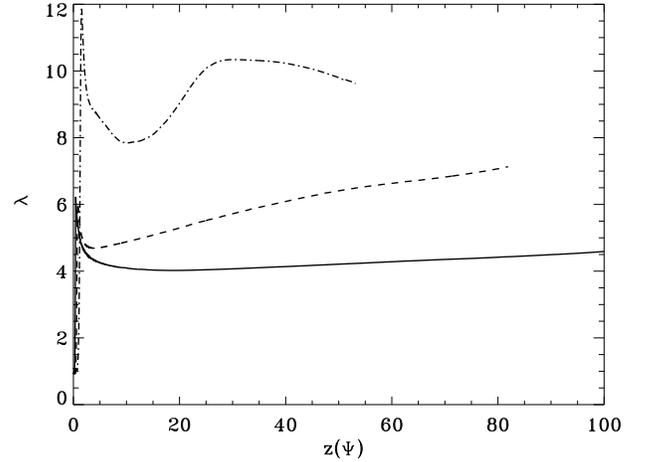}}
\caption{Magnetic lever arm parameter calculated along the fieldlines whose footpoints are located at  $r_0$ = 2  ({\it solid line}), 
             4 ({\it dashed line}) and 8 ({\it dot-dashed line}).}
\label{fig:lambda}
\end{figure}

\begin{figure}
\resizebox{\hsize}{!}{\includegraphics{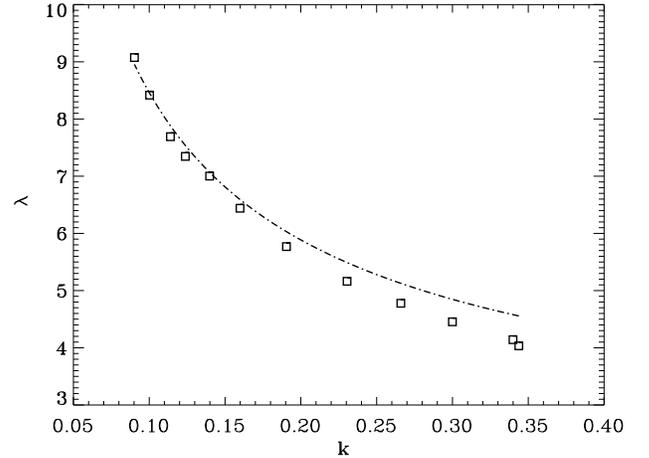}}
\caption{$k$-$\lambda$ relation ({\it squares}). The values are calculated at the Alfven point along different fieldlines anchored in the launching
            region of the disk ($1<r<10$). The value of $k$ increases going from the outer to the inner fieldlines. For comparison it is
            also plotted the relation Eq. (\ref{eq:weber}) found by Weber \& Davis (\cite{WebDav}) for a radial wind geometry ({\it dot-dashed line}).}
\label{fig:kaplam}
\end{figure}

The mass loading parameter $k$ is plotted in Fig.\ref{fig:kappa} while the $\lambda$ invariant is shown in Fig. \ref{fig:lambda}.
The two quantities show a much more constant behavior along the inner field line ({\it solid line}), indicating that the central part 
of the outflow has reached a more stationary state, while the outer part of it is still slowly evolving.
Anyway a clear trend emerges in these plots: the mass loading decreases going from the inner to the outer part of the outflow
while the magnetic lever arm increases. The relation between the $k$ and $\lambda$ parameters is shown in Fig. \ref{fig:kaplam}
where we plotted the values of these quantities calculated at the Alfven point on different fieldlines anchored in the disk.
The higher values of $k$ are found on the inner ($r\sim1$) fieldlines while the lower one in the outer part ($r\sim10$) of the launching
region. For comparison it is also plotted the relation:
\begin{equation} 
\lambda = \frac{3}{2}\left(1+k^{-2/3}\right)
\label{eq:weber}
\end{equation}
found by Weber \& Davis (\cite{WebDav}) for a radial wind geometry. The rough behavior $\lambda \propto k^{-2/3}$ is found also by 
Ouyed \& Pudritz (\cite{Ouyed}) and Anderson et al. (\cite{Anders04}).

We point out that in cold (adiabatic) steady solutions the following relation is valid 
(see e.g. Blandford \& Payne \cite{BlPayne82}) along each field line:
\begin{equation}
\frac{\dot{E}_\mathrm{jet, mag}}{\dot{E}_\mathrm{jet, kin}} = 2\frac{\dot{J}_\mathrm{jet, mag}}{\dot{J}_\mathrm{acc, kin}} = 2(\lambda -1)
\label{eq:lamratio}
\end{equation}
We plotted an averaged value over the disk surface of these two ratios in Fig. \ref{fig:mombal} and \ref{fig:enerbal}.
The magnetic lever arm is therefore a measure of the efficiency of the magneto-centrifugal mechanism: assuming that all the Poynting
flux available at the disk surface is completely converted into poloidal kinetic energy, the following relation for the poloidal asymptotic speed of the
outflow $u_{\mathrm{p},\infty}$ holds:
\begin{equation}
u_{\mathrm{p},\infty} = \Omega r_0 \sqrt{2\lambda-3}
\label{eq:uinf}
\end{equation}
Therefore the poloidal speed, expressed in units of the rotation speed at the footpoint, asimptotically assumes 
higher values on the outer field lines.

Finally, it is interesting to compare our solution with the cold self-similar steady solutions found by Casse \& Ferreira (\cite{CasFe00a})
characterized by the same parameters $\alpha_\mathrm{m}$, $\chi_\mathrm{m}$, $\epsilon$ and $\mu$.
The simulation presented in this section shows greater values of the mass loading $k$ and of the ejection parameter $\xi\sim0.1$
(calculated in Section \ref{sec:mrates}) and lower magnetic lever arms. The steady solution, characterized approximatively by
$k \sim 2\times10^{-2}$, $\xi \sim 10^{-2}$ and $\lambda\sim 35$, has therefore a much lower outflow rate and higher terminal speed.

A possible physical explanation is given by the shape of the current circuits shown in Section \ref{sec:currents}. Ferreira (\cite{Ferreira1997})
has shown that high ejection efficiencies ($\xi>0.5$) are obtained if the poloidal current enters the disk at its surface and that no steady
solution crossing all the critical points is possible. Its solutions are always characterized by a lower ejection parameter $\xi <0.5$ and by
a current outflowing from the disk surface. In our solutions the current returns and enters the disk in small region $1<r<2$ near the central
sink region: due to the internal boundary itself the rotation of the outflow goes suddenly to zero at the inner edge of the jet, creating a current sheet
which is forced to flow back inside the disk. This is therefore a numerical effect and a proper treatment of the region of interaction between the
disk and the central object is required to understand how the currents close in the inner region of the system.
Anyway it is possible to speculate that the returning current increases the ejection efficiency as proposed by Ferreira (\cite{Ferreira1997}): this
could be also confirmed by the fact that in our solution the mass loading parameter $k$ increases towards the center, where the current flows
back into the disk.

Another possible explanation is determined by the numerical dissipation which can increase the thermal energy and the pressure 
in the launching region, increasing the outflow rate as shown in Section \ref{sec:forces} (see also Casse \& Ferreira \cite{CasFe00b}).
If on one hand we showed that a low resolution yields the effects of a high disk diffusivity, on the other hand we cannot be confident 
that the standard resolution used in the simulations is sufficient to accurately resolve the vertical structure of the accretion disk:
it is well known that it is difficult for the type of
algorithms as the one used in this paper to handle low densities, as in the launching region, where the disk expands a lot. This is also
suggested by the behavior of the invariants $k$ and $\lambda$, e.g. on the inner field line considered: they show a slow transition to 
their more or less constant values, indicating that the transition between the resistive and the ideal MHD regimes happens on a scale
larger than the one determined by the vertical profile of the diffusivity (Eq. (\ref{eq:mdiff})).

\section{Effects of Ohmic heating}
\label{sec:joule}

\begin{figure}
\resizebox{\hsize}{!}{\includegraphics{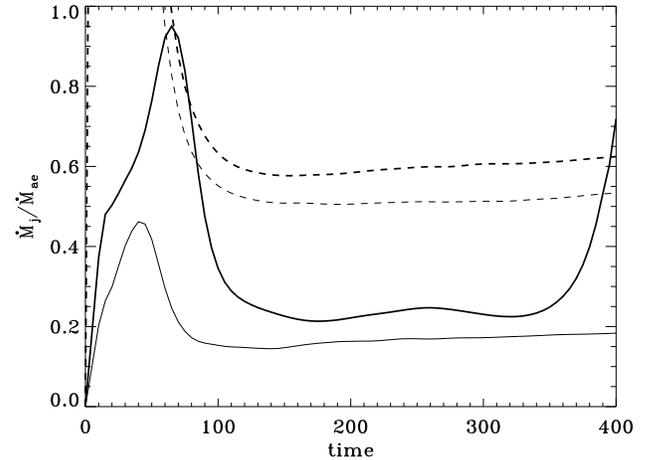}}
\caption{Temporal evolution of the ejection efficiency for the simulations in which the magnetic energy is 
             dissipated locally as Joule heating: ($\alpha_\mathrm{m}=1$, 
             $\chi_\mathrm{m}=3$, {\it solid thick line}) and $\alpha_\mathrm{m}=0.1$, $\chi_\mathrm{m}=1$,
             {\it dashed thick line}).  With a thinner line we also plot the curves which refer to the corresponding 
             cases in which the dissipated energy is radiated.}
\label{fig:mdissflux}
\end{figure}

We finally make a few considerations about the two simulations performed in which all the dissipated magnetic energy is released locally
inside the disk ($f=0$) and we will make a comparison with the corresponding cases in which the dissipated energy is radiated away ($f=1$).
We recall that these two simulations are characterized by ($\alpha_\mathrm{m}=0.1$, $\chi_\mathrm{m}=1$) and ($\alpha_\mathrm{m}=1$, 
$\chi_\mathrm{m}=3$).

The dissipative Ohmic term $\Lambda = \bar{\bar{\vec{\eta}}}\vec{J}\cdot\vec{J}$ acts now to increase the thermal energy and the entropy of the plasma 
inside the disk.
This has two important consequences: the energy dissipation at the midplane changes the disk thickness and modifies its structure while an increase of 
the pressure gradient at the disk surface allows more matter to be loaded in the outflow.

In Fig. \ref{fig:mdissflux} we plot the time evolution of the ejection efficiency of the two ``hot'' solutions and of the two corresponding ``cold'' 
simulations:  both the ``hot'' simulations are characterized by a higher efficiency; moreover, the case with $\alpha_\mathrm{m}=1$
({\it solid line}) shows at the end of the simulation a steep increase. This is due both to a decrease of the outer accretion rate and to an increase
of the outflow rate: the energy dissipation steepens the radial gradient of the thermal pressure inside the disk, thus slowing down the outer accretion; at the inner
radius the gradient is so steep that the disk does not accrete anymore and all the matter is pushed by the pressure gradient in a highly unsteady
outflow.

The disk thermal energy of the ``cold''  cases is almost constant in time. The  ``hot'' case with $\alpha_\mathrm{m}=0.1$ shows a similar behavior, 
with a disk thermal energy just slightly higher than the corresponding ``cold'' simulation. It is likely  that the higher outflow rate can balance the small increase
of the disk thermal energy which is equivalent to the $2\%$ of the accretion power that was ``radiated'' in the corresponding ``cold'' simulation (see Section \ref{sec:enerbud}).
On the other hand the disk thermal energy of the ``hot'' counterpart of the $\alpha_\mathrm{m}=1$ simulation continuously increases, not balanced by
the energy extracted by the outflow, leading to the unsteady behavior shown in Fig. \ref{fig:mdissflux}.
It is important to notice that the behavior of this simulation depends a lot on the expression that we used for the magnetic diffusivity (Eq. (\ref{eq:mdiff})) 
which is proportional to the disk thermal height scale: the Ohmic heating determines in fact an increase of the disk thickness that in turn implies a higher magnetic diffusivity.


\section{Summary and conclusions}

Our calculations have followed the long-term evolution of an
axisymmetric quasi-Keplerian magnetized disk 
up to the establishment of an inflow/outflow configuration. 
The accretion flow is driven by extraction of angular momentum of the
disk by the jet; this is shown by reaching balance between the
accretion and jet torques. The magnetic torque of the jet is most
efficient close to the surface of the disk extracting angular momentum
from the accretion flow; it stores angular momentum in the toroidal
magnetic field that then accelerates the outflowing plasma.
Therefore the simulations have succeeded in demonstrating 
that the magnetocentrifugal mechanism originally proposed by 
Blandford \& Payne can launch jets, provided certain physical conditions 
on the magnetic resistivity and initial field configuration are satisfied.

In particular we have shown that an isotropic ($\chi_\mathrm{m}=1$) resistive
configuration with $\alpha_\mathrm{m}=0.1$, due to the stronger advection of
the field compared to its diffusion, produces highly unsteady magnetic 
structures, like it was displayed in Fig. \ref{fig:A01T1CT} and in the left panel of
Fig. \ref{fig:imagebar}. This is in agreement with the stationary models of
Casse \& Ferreira (\cite{CasFe00a}) according to which it is not possible to obtain a steady outflow
with such a low $\alpha_\mathrm{m}$ parameter except for highly anisotropic
configurations ($\chi_\mathrm{m} > 10^2$). This would be in agreement with the
results shown in Section \ref{sec:jrates}: a much stronger toroidal resistivity should
reduce the torque exerted by the jet on the disk and thereby its accretion rate 
(Eq. (\ref{eq:acctor})). 
On the other hand we have also shown how this unsteady behavior depends strongly
on the resolution assumed and therefore on the numerical dissipation: 
the same case $\alpha_\mathrm{m}=0.1$ performed with a resolution four times lower
presents many of the characteristics of a solution with $\alpha_\mathrm{m}=1$.
It can be presumed that the quasi-stationary behavior found by Casse \& Keppens 
(\cite{CasKe02}, \cite{CasKe04}) with an isotropic $\alpha_\mathrm{m}=0.1$ parameter
can be affected by the resolution used.

On the other hand, also the cases characterized by $\alpha_\mathrm{m}=1$ are not 
perfectly stationary, despite of showing an ordered magnetic configuration favorable
for a steady launching: in all the simulations performed some integrated quantities, like
the jet torque (Fig. \ref{fig:mombal}) or the accretion energy (Fig. \ref{fig:enerbal}), 
are still slowly evolving. Even the case characterized by an anisotropic ($\chi_\mathrm{m}=3$)
diffusivity, needed by the stationary models, does not show a perfectly stationary behavior; 
moreover it can be characterized by adimensional quantities like $\xi$, $k$ or $\lambda$ which
differ a lot from the analogous found in the cold solutions of Casse \& Ferreira (\cite{CasFe00a}). 
In Section \ref{sec:statio} we proposed some physical and numerical reasons to explain this behavior.

Nevertheless the constant value assumed in time by some quantities, like the ejection efficiency (see Fig. 
\ref{fig:mflux}) or the ratio between the accretion power and the jet energy flux (central panel of Fig.
\ref{fig:enerbal}) suggests that our solutions are slowly evolving through a series of quasi-stationary
states. Some clear trend emerges: increasing the poloidal and/or the toroidal resistivity the ejection efficiency
decreases, going from 55\% in the ($\alpha_\mathrm{m}=0.1$, $\chi_\mathrm{m}=1$) case to 20\%
for the  ($\alpha_\mathrm{m}=1$, $\chi_\mathrm{m}=3$) simulation.
From the energetic point of view, our simulations show that more than 90\% of the energy liberated in the 
accretion inflow is released in the jet and that less dissipative cases produce slightly more powerful jets. 
Correspondingly, the radiated power that must take care of the Joule heating dissipation is less than 10\%.
In Section \ref{sec:joule} we have also shown that in the more dissipative case with $\alpha_\mathrm{m}=1$,
if this energy is released inside the disk instead of being radiated, the accretion flow is disrupted and a highly 
unsteady outflow is formed.

We have also demonstrated how the efficiency of the magneto-centrifugal mechanism, as measured by the
ratio of the Poynting flux to the kinetic flux at the disk surface, is affected by the resistive configuration. 
A higher value of the poloidal and/or toroidal resistivity determines a  greater magnetic lever arm, 
linked to the energy flux ratio as stated by Eq. (\ref{eq:lamratio}). Our simulations show also that the
$\lambda$ parameter, as measured by Eq. (\ref{eq:lamratio}), is related with a good approximation to the
ejection parameter $\xi$ (in the cases where this parameter has been measured):
\begin{equation}  
\lambda \sim 1+\frac{1}{2\xi}
\end{equation}
as stated by the self-similar models by Ferreira (\cite{Ferreira1997}).

We finally try to test if observations of outflows from classical T-Tauri stars can 
give some constraint on the parameters of our simulations.
The size of the launching region measured in our simulations (from 0.1 to 1 AU)
is consistent with the estimates derived from observations (Bacciotti et al. \cite{Bacciotti}).
On the other hand, the one-sided ejection efficiency inferred from the observations 
gives values, even with high uncertainty, in the range 0.01-1 (Cabrit \cite{Cabrit}): this result
seems to favor our solution characterized by ($\alpha_\mathrm{m}=1$, $\chi_\mathrm{m}=3$).
Even if, in agreement with observations, all our solutions are characterized by a speed at the upper boundary 
of the computational domain which is a few times the escape speed from the potential well of the central 
object (Fig. \ref{fig:velz}), a quantitative comparison with the poloidal speed (Bacciotti et al. \cite{Bacciotti}) and rotation 
signatures (Bacciotti et al. \cite{Bacciotti2002}, Coffey et al. \cite{Coffey2004}, Woitas et al. \cite{Woitas2005})
of the observed outflows is much more difficult: the size of our computational domain, which reaches a physical scale along $z$ 
equal to 12 AU, is around three times smaller than the distance from the source investigated by current HST observations
($>$ 30 AU). Anyway, as it has been pointed out by Ferreira et al. (\cite{Ferreira2006}), a solution characterized by a magnetic lever arm 
$\lambda\sim 10$ successfully reproduces  the values of both the poloidal and toroidal speeds at the 
currently observed spatial scale: the highest lever arm value found in our simulations ($\lambda\sim9$) is observed in the outer part 
of the launching region of the case ($\alpha_\mathrm{m}=1$, $\chi_\mathrm{m}=3$) (Fig. \ref{fig:kaplam}).

Two specific conclusions and difficulties of our calculations must be
mentioned. First, the magnetocentrifugal process appears to operate
only with relatively strong magnetic fields in the disk; a strong
field tends to inhibit the development of turbulence that could be the
origin of magnetic resistivity. Instabilities (shear, Kelvin-Helmholtz) 
other than magneto-rotational might produce the required turbulence. 
On the other hand,  the resistivity inside a protostellar disk can be 
extremely high between 0.1 and 10 AU, where the grains become the dominant
charge carriers (Wardle \cite{Wardle}). The coupling between the accreting plasma
and the magnetic field is therefore strongly, perhaps too much, reduced.
The second point is that outflow rates measured in our simulations 
turn out to be rather high, which applies well to jets in star
formation regions, where extended bipolar structures are observed; for
AGN nuclei one must go to the relativistic limit (which we have not
done here), and also discuss how the (possibly subsonic) wind from the
disk outer part interacts with the surrounding galaxy.

\begin{acknowledgements}

We thank Timur Linde for help in upgrading the MHD version of the
FLASH code. CZ would also thank Jonathan Ferreira, Zakaria Meliani
and Christophe Sauty for all the valuable suggestions and discussions.
The simulations have been performed at IDRIS (Orsay) under the Project 
HPC-EUROPA (RII3-CT-2003-506079), with the support of the European 
Community - Research Infrastructure Action under the FP6 ``Structuring the
European Research Area''  Program.
The authors acknowledge support through the Marie Curie Research Training 
Network JETSET (Jet Simulations, Experiments and Theory) under contract
MRTN-CT-2004-005592.
This work has been supported in part by the
U.S. Department of Energy under grant No. B523820 to the Center of
Astrophysical Thermonuclear Flashes at the University of Chicago and
by the Italian Ministero dell'Universit\`a e della Ricerca under a
contract PRIN 2005 to the University of Torino. 
The software used in this work was in part developed by the 
DOE-supported ASC / Alliance Center for Astrophysical Thermonuclear 
Flashes at the University of Chicago.

\end{acknowledgements}

\end{document}